\documentclass[
 reprint,
superscriptaddress,
 amsmath,amssymb,
 aps,
]{revtex4-2}

\usepackage{graphicx}
\usepackage{dcolumn}
\usepackage{bm}
\usepackage{tabularx}
\usepackage{xcolor}

\begin{document}

\preprint{APS/123-QED}

\title{Towards a fictitious magnetic field trap\\for both ground  and Rydberg state $^{87}$Rb atoms via the evanescent field of an optical nanofibre}

\author{Alexey Vylegzhanin}
\email{alexey.vylegzhanin@oist.jp}
\altaffiliation{
author to whom any correspondence should be addressed.}
\affiliation{Light-Matter Interactions for Quantum Technologies Unit, Okinawa Institute of Science and Technology Graduate University, Onna, Okinawa 904-0495, Japan.}

\author{Dylan~J.~Brown}
\altaffiliation[Current address: ]{Centre for Cold Matter, Blackett Laboratory, Imperial College London, Prince Consort Road, London, SW7 2AZ UK}
\affiliation{Light-Matter Interactions for Quantum Technologies Unit, Okinawa Institute of Science and Technology Graduate University, Onna, Okinawa 904-0495, Japan.}

\author{Danil~F.~Kornovan}
\affiliation{Center for Complex Quantum Systems, Department of Physics and Astronomy, Aarhus University, Ny Munkegade 120, DK-8000 Aarhus C, Denmark.}

\author{Etienne Brion}
\affiliation{Laboratoire Collisions Agr\'{e}gats R\'{e}activit\'{e}, FeRMI, Universit\'{e} de Toulouse and CNRS UMR 5589, Toulouse, France.}

\author{S\'{i}le~{Nic Chormaic}}
\email{sile.nicchormaic@oist.jp}
\altaffiliation{
author to whom any correspondence should be addressed.}
\affiliation{Light-Matter Interactions for Quantum Technologies Unit, Okinawa Institute of Science and Technology Graduate University, Onna, Okinawa 904-0495, Japan.}

\date{\today}

\begin{abstract}
Cold Rydberg atoms, known for their long lifetimes and strong dipole-dipole interactions that lead to the Rydberg blockade phenomenon, are among the most promising platforms for quantum simulations, quantum computation and quantum networks. However, a major limitation to the performance of Rydberg atom-based platforms is dephasing, which can be caused by atomic motion within the trap. Here, we propose a trap for $^{87}$Rb cold atoms that confines both the electronic ground state and a Rydberg state, engineered to minimize the differential light shifts between the two states. This is achieved by combining a fictitious magnetic field induced by optical nanofibre guided light and an external bias magnetic field. We calculate trap potentials for the cases of one- and two-guided modes with quasi-linear and quasi-circular polarisations, and calculate trap depths and trap frequencies for different values of laser power and bias fields.  Moreover, we discuss the impact of the quadrupole polarisability of the Rydberg atoms on the trap potential and demonstrate how the size of a Rydberg atom influences the ponderomotive potential generated by the nanofibre-guided light field. This work expands on the idea of light-induced fictitious magnetic field traps and presents a practical approach for creating quantum networks using Rydberg atoms integrated with optical nanofibres to generate 1D atom arrays.

\end{abstract}

\maketitle


\section{Introduction}
Cold Rydberg atoms are a promising platform for quantum information~\cite{demonstration_of_C_NOT_gate, quantum_information_Saffman, adams2019rydberg_quantum_tech, jaksch2000fast_gates_Rydberg, PhysRevResearch.5.013205, chew2022ultrafast} and quantum simulation~\cite{weimer2010rydberg_simulator, scholl2021quantum, bharti2022ultrafast} due to the long lifetimes of the excited states and the strong dipole-dipole interaction resulting in Rydberg blockade~\cite{Urban2009}. The Rydberg blockade allows for the deterministic entanglement of qubits~\cite{Rydberg_Bell_states}, implementation of C-NOT and C-Phase quantum gates~\cite{demonstration_of_C_NOT_gate,one_pulse_CPhase}, and the generation of single-photon emitters~\cite{ripka2018room_single_photon} and single-photon switches~\cite{baur2014single_switch}. Typically, Rydberg experiments are performed in free-space, often with atoms being excited to the desired Rydberg state in optical tweezers arrays~\cite{Rydberg_in_tweezers,barredo2020three}, optical lattices~\cite{Rydberg_in_optical_lattices} or micron-sized vapour cells~\cite{mum_sized_cell}. Emerging platforms include hybrid systems such as atom-waveguide configurations ~\cite{hollowfiber2014rydberg,Ke:16,PhysRevA.102.063703, jones2020collectively, li2024atom, vylegzhanin2024light}, and, specifically, Rydberg-waveguide systems~\cite{KP_rydberg_generation, Vylegzhanin:23, zhang2024chiral,ocola2024control}, atom chips~\cite{de2018coherent}, and Rydberg-cavity setups~\cite{cavity_Rydberg}. Compared to free-space systems, such  platforms have advantages, e.g. low power consumption and high scalability~\cite{quantum_information_Saffman, sunami2024scalable}, which are important for creating practical quantum devices. 

In order to perform experiments with Rydberg atoms next to devices, such as waveguides or atom chips~\cite{de2018coherent, schmied2010optimized}, one needs to trap both the ground and Rydberg state atoms in close proximity to the device itself or one must briefly turn the trap off such that the atoms are essentially frozen during the measurement process,  as done in optical tweezers experiments~\cite{gaetan2009observation, chew2022ultrafast}. Keeping the atoms as stationary as possible during the experimental sequence is critical so they do not experience unwanted dephasing when excited to the Rydberg state~\cite{saffman2011rydberg}. There are already   several proposed and demonstrated methods to trap Rydberg atoms in tight magnetic microtraps ~\cite{boetes2018trapping,lesanovsky2005magnetic}, Ioffe-Pritchard traps~\cite{hezel2007ultracold}, by Z-wires ~\cite{anderson2013production}, bottle beam traps~\cite{zhang2011magic}, optical lattices using a ponderomotive potential~\cite{topcu2013intensity}, and  Laguerre-Gaussian  beams~\cite{cortinas2020laser}. Another approach to Rydberg-state quantum technologies uses alkaline-earth atoms. With two valence electrons, they can be trapped in optical tweezers similarly to ground-state atoms \cite{wilson2022trapping}.

 Here, we propose an optical nanofibre (ONF)- based trap for $^{87}\mathrm{Rb}$ atoms in both the ground 5S$_{1/2}$ and select Rydberg  nD$_{5/2}$ states formed by combining an offset magnetic field with a light-induced fictitious magnetic field. The fictitious magnetic field is experienced by atoms in a light field which has a non-zero ellipticity of the electric field~\cite{cohen1972experimental,le2013dynamical}. Earlier work by Schneeweiss \textit{et al.}~\cite{schneeweiss2014nanofiber} proposed a similar type of magnetic trap for cold $^{133}$Cs atoms near an ONF, but this was limited to ground state atoms.  While our approach could be viewed as analogous to works on trapping Rydberg atoms in magnetic microtraps~\cite{boetes2018trapping} or next to Z-shaped wires~\cite{anderson2013production}, our proposed scheme benefits from the presence of the ONF for trapping the atoms, thereby facilitating efficient coupling of photons emitted by the atoms into the waveguide that could act as a link between quantum nodes~\cite{li2024atom, zhang2024chiral}. One of the most promising applications of Rydberg states near waveguides is in all-fiber quantum networks, where they can serve as nanofiber-based quantum repeater nodes, following the scheme introduced in~\cite{zhao2010efficient}, enabling entanglement distribution by straightforward integration with optical fiber links~\cite{sunami2024scalable}.

The paper is organised as follows. In Sec. II we describe the concept of the light-induced fictitious magnetic field. In Sec. III we derive the equation for the combined magnetic field trap and calculate trapping potentials for the $49$D$_{5/2}$ Rydberg state. In Sec. IV we determine particular Rydberg states for which the trap potential is comparable to that of the 5S$_{1/2}$ ground state in order to minimize any dephasing from different trap potentials. In Sec. V we discuss how the size of the Rydberg atom could affect the trap potential due to changes in the electromagnetic field across its dimensions~\cite{zhang2011magic} and also the energy shift introduced by the quadrupole AC Stark shift. Our conclusions are given in Sec. VI. While the proposed concept may be experimentally challenging, we have previously shown that Rydberg state atoms can be prepared next to an ONF ~\cite{KP_rydberg_generation,Vylegzhanin:23} from 5S$_{1/2}$ ground state atoms so the scheme would seem to be technically feasible. Note that all wavelengths, $\lambda$, in the following are vacuum wavelengths.

\section{Light-induced fictitious magnetic fields}
An atom in an oscillating electromagnetic field experiences an AC-Stark shift dependent on the magnitude of the electric field and the frequency-dependent atom polarisability, which can be represented by the scalar, vector, and tensor components of the polarisability tensor~\cite{le2013dynamical}. Scalar light shifts can be tuned close to zero by using a tune-out wavelength, $\lambda_{\mathrm{to}}$, so that only the vector and tensor light shifts remain. For the $^{87}\mathrm{Rb}$ ground state, $\mathrm{5S}_{1/2}$, the tensor light shift is zero and $\lambda_{\mathrm{to}} \approx 790.2$~nm~\cite{leblanc2007species}. At this wavelength, the scalar light shifts from the D$_1$ and D$_2$ transitions are equal and opposite in sign for the $\mathrm{5S}_{1/2}$ state and result in a net zero scalar shift of this state. Therefore, the only component remaining is the vector light shift, which depends on the magnetic quantum number, $m_F$, and can be written as the effect of a light-induced fictitious magnetic field given by

\begin{equation} \label{eq:1}
    \boldsymbol{B}_{\mathrm{fict}_F}=\frac{\alpha^{\mathrm{v}}_{nJF}}{8\mu_B g_{nJF} F}i[\boldsymbol{\mathcal{E}}^*\times \boldsymbol{\mathcal{E}}],
\end{equation}
where $\mu_B$ is the Bohr magneton, $n$ is the principle quantum number, $J$ is the total angular momentum quantum number, $F$ is the hyperfine splitting quantum number, $g_{nJF}$ is the Landé g-factor, $\alpha^{\mathrm{v}}_{nJF}$ the frequency-dependent vector polarisability of the $\left|nJF\right>$ state, and $\boldsymbol{\mathcal{E}}$ is the positive-frequency electric field envelope of the complex electric field $\mathbf{E}=1/2\left(\boldsymbol{\mathcal{E}}e^{-i\omega t} + \mathrm{c.c.}\right)$~\cite{yang2008optically, schneeweiss2014nanofiber}. 

\begin{figure*}
    \centering
    \includegraphics[width=12cm]{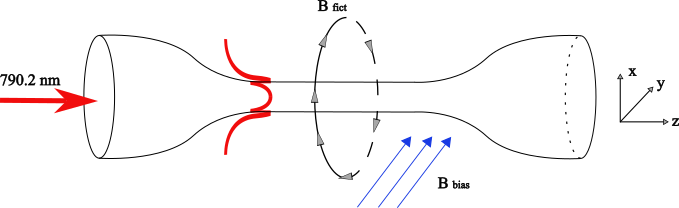}
    \caption{Schematic of the optical nanofibre magnetic trap.  The ONF-guided fundamental mode with a  wavelength $\lambda = 790.2$~nm is indicated by the red curve. The resulting fictitious magnetic 
    field, $B_{\mathrm{fict}}$, surrounds the fibre (grey arrows) and is added to a bias magnetic field, $B_{\mathrm{bias}}$ (blue arrows) to form the total trap potential, $U_\mathrm{tot}$.}
    \label{fig:scheme}
\end{figure*}

\begin{figure}[t]

    \raggedright{a)}
    
    \centering
    \includegraphics[width=8cm]{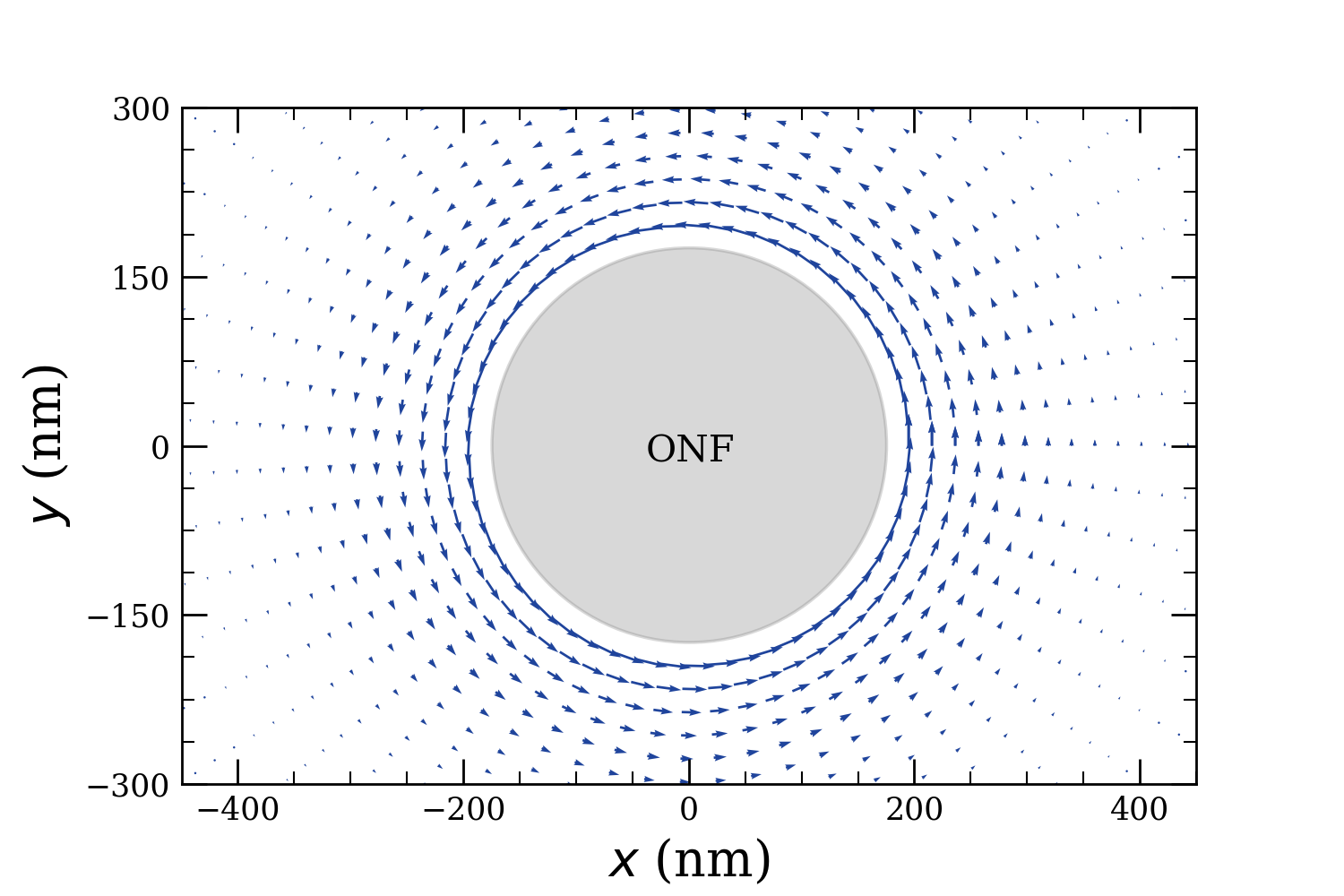}

    \raggedright{b)}

    \centering
    \includegraphics[width=8cm]{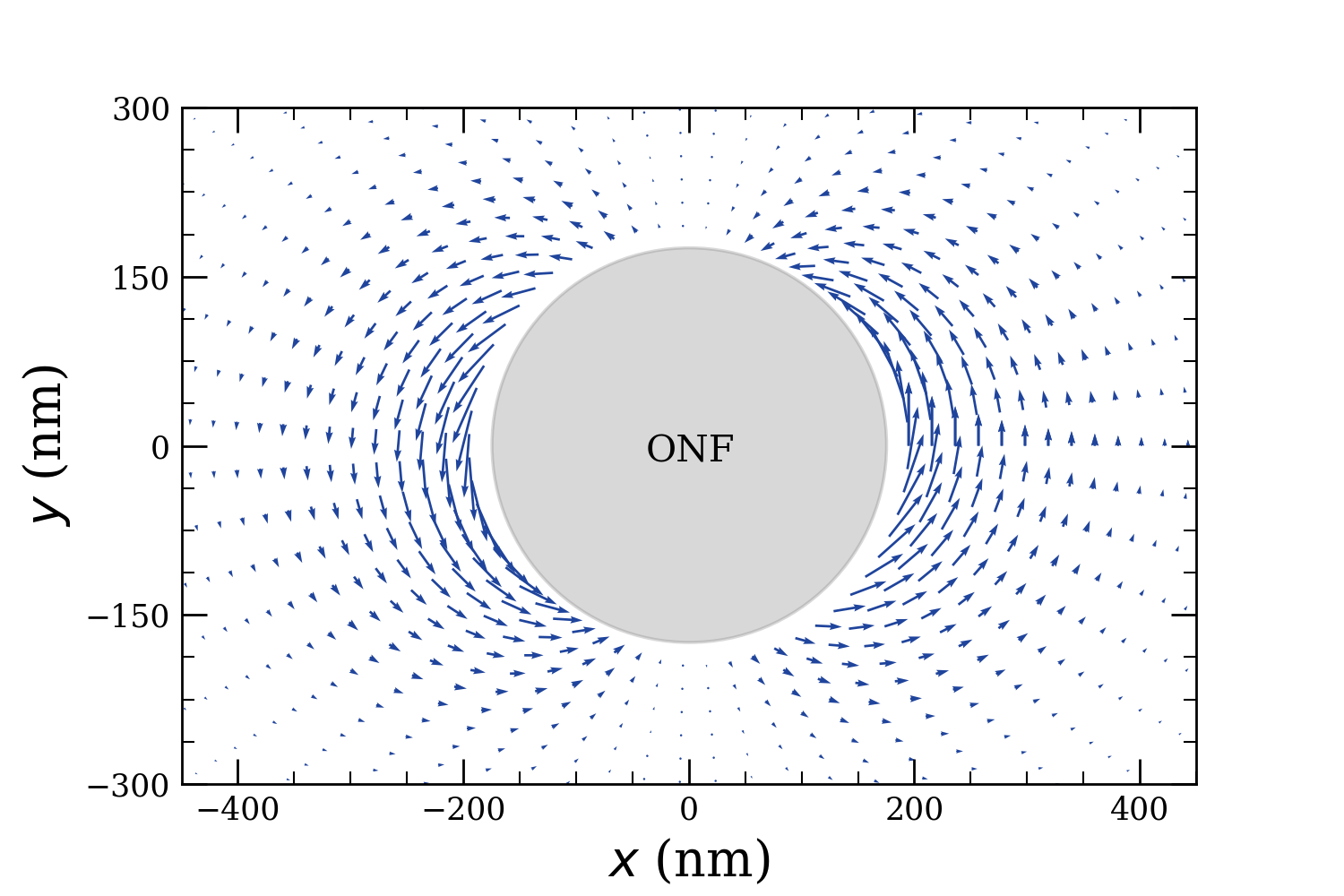}
    \caption{Normalised vector field of the light-induced fictitious magnetic field (blue arrows) for a $^{87}$Rb atom in the $\mathrm{\left|49D_{5/2}, m_J=5/2\right>}$ state in the $xy-$plane perpendicular to the fibre axis for (a)  quasi-circularly and (b)  quasi-linearly polarised fundamental guided modes of the optical nanofibre. The fibre radius is $a=175$~nm and the  wavelength is $\lambda=790.2$~nm.}
    \label{magnetic_vector_field}
\end{figure}

Choosing particular Rydberg states with scalar and tensor polarisabilities close to zero, we can obtain a similar scenario when only the vector shift is present. For Rydberg states of high principal quantum number, n, the hyperfine coupling between the electron spin and the nuclear spin is negligible, so it is more natural to consider the angular momentum, $J$, and its projections, $m_J$. One can rewrite the light-induced fictitious magnetic field in the following form~\cite{le2013dynamical}
\begin{equation} \label{eq:2}
    \boldsymbol{B}_{\mathrm{fict}_J}=\frac{\alpha^{\mathrm{v}}_{nJ}}{8\mu_B g_{nJ} J}i[\boldsymbol{\mathcal{E}}^*\times \boldsymbol{\mathcal{E}}], 
\end{equation}
where   $g_{nJ}$ is the Landé g-factor and $\alpha^{\mathrm{v}}_{nJ}$ is the vector polarisability for the $\left|nJm_J\right>$ Rydberg state with $m_J$ being the magnetic quantum number. One can notice that the direction of the induced fictitious magnetic field depends on the signs of $\alpha^{\mathrm{v}}_{nJ}$ and $g_{nJ}$ as well as the cross-product of the electric field. The cross-product is non-zero only if the electromagnetic field has elliptical polarisation; significantly, this is the case for both quasi-linearly (QL) and quasi-circularly (QC) polarised fundamental guided modes of an optical nanofibre~\cite{le2004field}.

\section{Fictitious magnetic field trap}
To model the trap, we consider an optical nanofibre made of silica with refractive index $n=1.44$ and radius $a=175$~nm. For this radius, the fibre only supports the HE$_{11}$ fundamental mode for all wavelengths used in this work~\cite{frawley2012higher, le2017higher}.  We define the light propagation direction to be along the $z-$axis. A significant portion of the guided light exists in the evanescent field outside the optical nanofibre~\cite{le2004field}. This evanescent field can be used, for example, to probe atoms in the vicinity of the fibre~\cite{nayak_first_fiber_probe, gokhroo2022rubidium} or to create a two-colour dipole trap for atoms~\cite{vetsch2010optical,gupta2022machine,liedl2022observation,pache2024realization} or molecules~\cite{marquez2023nanofibre}. A schematic of the setup is shown in Figure~\ref{fig:scheme}.

The light-induced fictitious magnetic field  presented in Eq.~\ref{eq:2} can be expressed in cylindrical coordinates as~\cite{le2013state}
\begin{equation}
    \boldsymbol{B}_{\mathrm{fict}_J}=\frac{\alpha^{\mathrm{v}}_{nJ}}{4\mu_B g_{nJ} J}[\mathrm{Im}(\mathcal{E}_z\mathcal{E}^*_r) {\hat{\boldsymbol{\phi}}}+\mathrm{Im}(\mathcal{E}_r\mathcal{E}^*_\phi){\hat{\boldsymbol{z}}}+\mathrm{Im}(\mathcal{E}_\phi\mathcal{E}^*_z){\hat{\boldsymbol{r}}}],
\end{equation}

\noindent where $\boldsymbol{\mathcal{E}}_\mathrm{circ}^{(fp)}=(\mathcal{E}_r, \mathcal{E}_{\phi}, \mathcal{E}_z)$ represents the radial, azimuthal, and longitudinal cylindrical components of the electric field of the nanofibre guided mode. 

Outside the ONF (i.e., $r>a$), the electric field components of the QC polarised fundamental guided mode $\mathrm{HE}_{11}$ are \cite{le2017higher}

\begin{equation}\label{eq:QC field}
\begin{split}
    & \mathcal{E}_r=iA\frac{\beta}{2q}\frac{J_1(ha)}{K_1(qa)}[(1-s)K_0(qr)+(1+s)K_2(qr)]e^{i\beta z + i p\phi}, \\
    & \mathcal{E}_{\phi}=-pA\frac{\beta}{2q}\frac{J_1(ha)}{K_1(qa)}[(1-s)K_0(qr)-(1+s)K_2(qr)]e^{i\beta z + i p\phi}, \\ 
    & \mathcal{E}_z=fA\frac{J_1(ha)}{K_1(qa)}K_1(qr)e^{i\beta z + ip \phi},
\end{split}
\end{equation}
 
\noindent where $A$ is a normalisation factor dependent on the power of the guided light, $p=\pm$ is the handedness of the circular polarisation, $f=\pm 1$ is the direction of propagation of the guided light and the parameters $s,h,$ and $q$ are given by
\begin{equation}
\begin{split}
    & s=\left( \frac{1}{h^2a^2}+\frac{1}{q^2a^2} \right)/\left(\frac{J_1'(ha)}{haJ_1(ha)}+\frac{K_1'(qa)}{qaK_1(qa}\right), \\
    & h=(n_1^2k^2-\beta^2)^{1/2},\\
    & q=(\beta^2-n^2_2k^2)^{1/2}.
\end{split}
\end{equation}

\noindent Here, $\beta$ is the propagation constant of the guided mode, $k=2\pi/\lambda$ is the vacuum wavenumber of the light field, and $K_n$ and $J_n$ denote the Bessel functions of the first kind and the modified Bessel functions of the second kind, respectively.  The fictitious magnetic vector field is shown in  Fig.~\ref{magnetic_vector_field}(a).

Similarly, for the QL mode, the electric field has non-zero ellipticity in the $zr-$plane. The electric field of a QL mode with polarisation angle $\phi_\mathrm{pol}$ with respect to the $x-$axis, see Fig.~\ref{fig:scheme}, can be written as a summation of fields with opposite circular polarisations

\begin{equation}
    \boldsymbol{\mathcal{E}}_{\mathrm{lin}}^{(f\phi_\mathrm{pol})}=\frac{1}{\sqrt{2}}(\boldsymbol{\mathcal{E}}_\mathrm{circ}^{(f+)}e^{-i\phi_\mathrm{pol}}+\boldsymbol{\mathcal{E}}_\mathrm{circ}^{(f-)}e^{i\phi_\mathrm{pol}}).
\end{equation}
.

\noindent For a QL field, the fictitious magnetic field (Eq.~\ref{eq:2}) can be simplified to~\cite{le2013state}

\begin{equation}\label{eq:QL_B}
\boldsymbol{B}_{\mathrm{fict}_J}=\frac{\alpha^{\mathrm{v}}_{nJ}}{4\mu_B g_{nJ} J}[\mathrm{Im}(\mathcal{E}_{z, \mathrm{lin}}\mathcal{E}^*_{r,\mathrm{lin}}) {\hat{\boldsymbol{\phi}}}+\mathrm{Im}(\mathcal{E}_{\phi, \mathrm{lin}}\mathcal{E}^*_{z,\mathrm{lin}}){\hat{\boldsymbol{r}}}].  
\end{equation}
The fictitious magnetic vector field for this case is shown in  Fig.~\ref{magnetic_vector_field}(b).

We now introduce a bias magnetic field, $\boldsymbol{B}_\mathrm{bias}$, to produce a region where the total effective magnetic field sums to zero and atoms may be trapped. The fictitious magnetic field, $\boldsymbol{B}_{\mathrm{fict}_J}$, behaves like a real magnetic field~\cite{le2013dynamical,schneeweiss2014nanofiber} and can be vector added to the bias field, as shown in earlier experimental works~\cite{cohen1972experimental, zielonkowski1998optically}. The total effective magnetic field is then simply $\boldsymbol{B}_{\mathrm{eff}}=\boldsymbol{B}_{\mathrm{fict}_J}+\boldsymbol{B}_\mathrm{bias}$.

\begin{figure}
    
    \raggedright {a)}
    
    \centering
    \includegraphics[width=8cm]{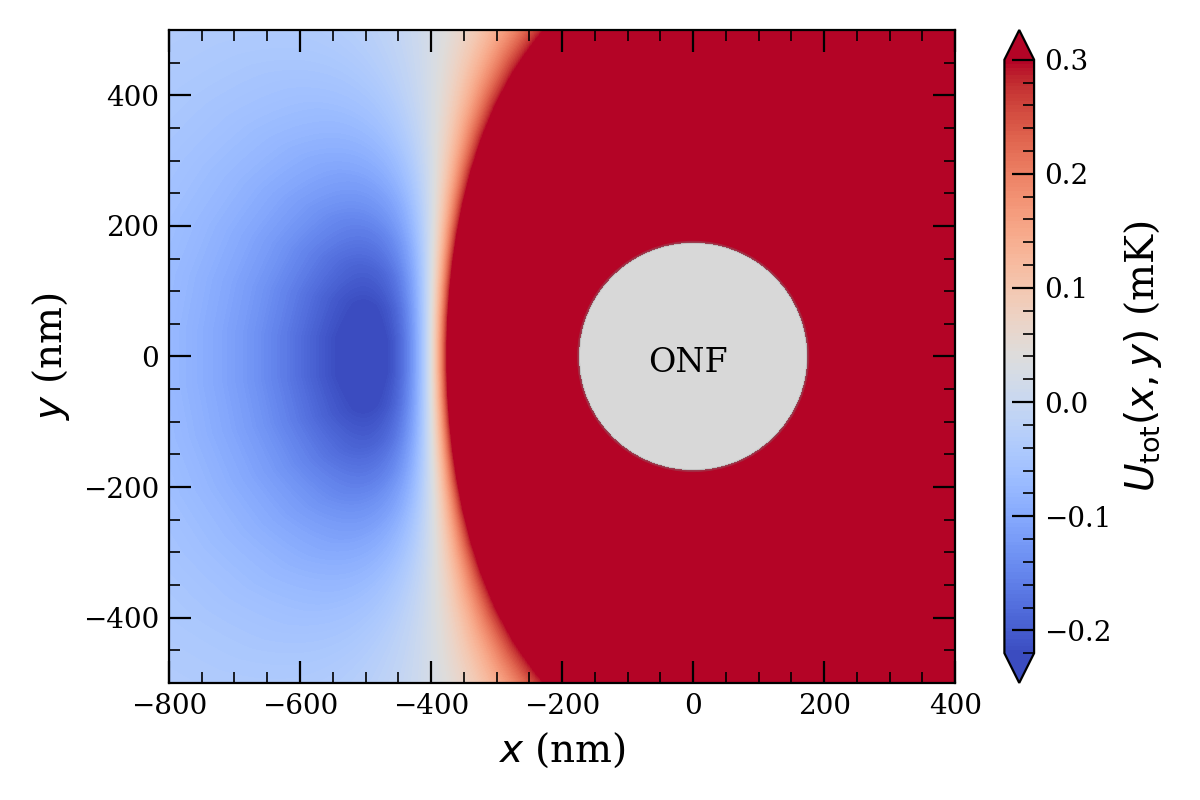}

    \raggedright{b)}

    \centering
    \includegraphics[width=8cm]{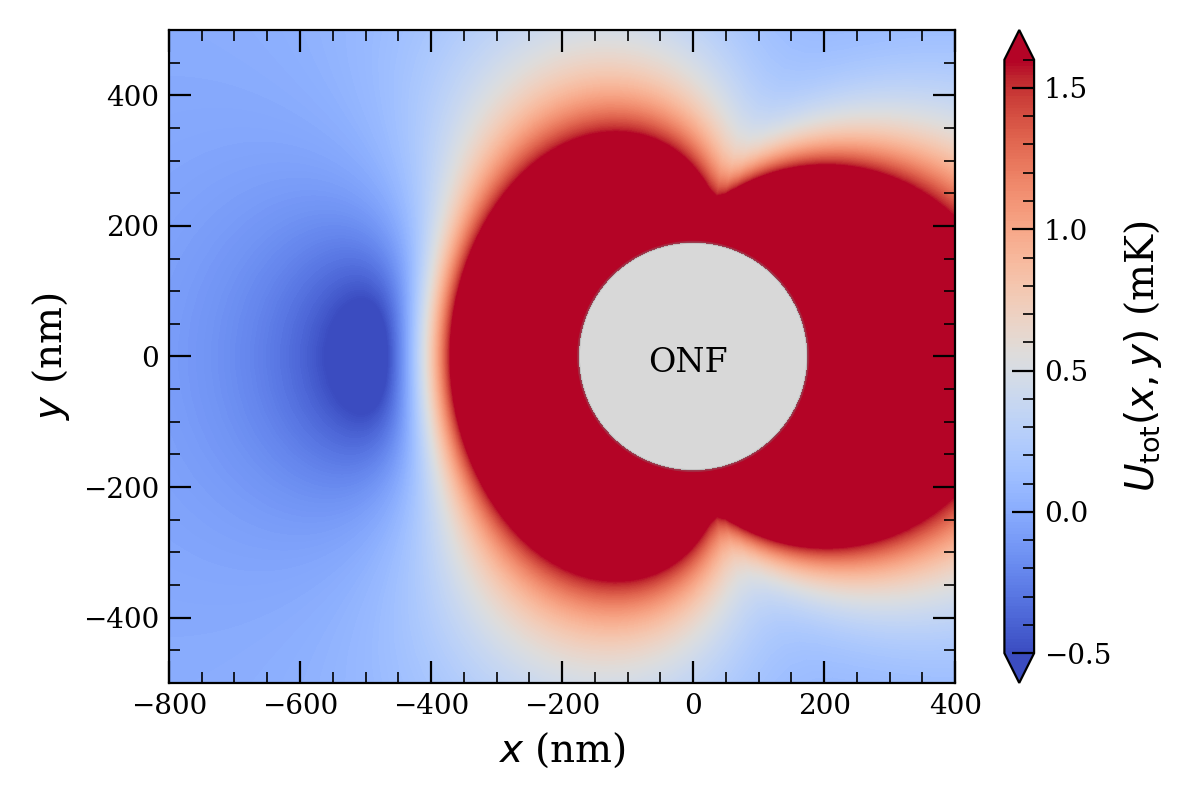}
    \caption{Two-dimensional plots of the fictitious magnetic trap potentials, $\mathrm{U_{tot}(x,y)}$, for a $^{87}$Rb atom in the $\mathrm{\left|49D_{5/2}, m_J=5/2\right>}$ Rydberg state. The potentials are formed by  (a) a quasi-circularly polarised mode with $P=20$~mW and $B_{\mathrm{bias}}=60$~G and (b) a quasi-linearly polarised mode with $P=10$~mW and $B_{\mathrm{bias}}=6$~G. The potential minima in both cases are formed on the left hand side of the fibre with potential depths on the order of 0.25~mK. The fibre radius is $a=175$~nm and the  wavelength is $\lambda=  790.2$~nm. }
    \label{fig:2d_trap}
\end{figure}

For both polarisation cases, we set the bias magnetic field perpendicular to the $z$-axis and in the $+y$ direction, see Fig.~\ref{fig:scheme}. The magnetic potential an atom experiences is 
\begin{equation}
    U_{\mathrm{mag}}=-\boldsymbol{\mu}\cdot\boldsymbol{B}_{\mathrm{eff}},
\end{equation}
where $\boldsymbol{\mu}$ is the magnetic moment of the atom~\cite{le2013dynamical}. The potential here is formed for low-field seeking atoms. If the magnetic moment of the atom remains anti-parallel to the local effective magnetic field during atomic motion in the trap, the trap potential can be simplified to

\begin{equation}\label{eq:8}
    U_{\mathrm{mag}}=\mu_Bg_{nJ}m_J|\boldsymbol{B}_{\mathrm{eff}}|,
\end{equation}
while the quantisation axis is set along the $z-$axis by the $z-$component of the light-induced fictitious magnetic field in the QC mode case, and by another external magnetic field along the $z-$axis in the QL mode case.

\subsection{Trapping Rydberg state atoms }
 
 We consider a $^{87}$Rb atom in the 49D$_{5/2}$ Rydberg state, which,  for the vacuum wavelength of the guided mode,  $\lambda=790.2$~nm, has a vector polarisability $\alpha_{nJ}^{\mathrm{v}}=-9.7\cdot 10^{-5}~\mathrm{Hz\cdot m^2/V^2}$ and a scalar polarisability $\alpha_{nJ}^{\mathrm{sc}}=-7.2\cdot 10^{-6}~\mathrm{Hz\cdot m^2/V^2}$, which were calculated using the Alkali-Rydberg-Calculator (ARC)~\cite{robertson2021arc}, and a ponderomotive polarisability $\alpha_{\mathrm{pd}}=-15\cdot 10^{-6}~\mathrm{Hz\cdot m^2/V^2}$. We 
 calculate the ponderomotive polarisability by $\alpha_{\mathrm{pd}}=-e^2/( m_\mathrm{e} \omega^2)$, where $-e$ is the electron charge, $m_{\mathrm{e}}$ is the electron mass, and $\omega$ is the trap light frequency, i.e. the frequency of the light propagating in the ONF.
 Since the scalar polarisability and the ponderomotive polarisability are only one order of magnitude less than the vector polarisability, the scalar energy shift, $U_{\mathrm{sc}}=-1/4\alpha_{nJF}^{\mathrm{sc}}|\boldsymbol{\mathcal{E}}|^2$, and the ponderomotive energy shift, $U_{\mathrm{pd}}=-1/4\alpha_{\mathrm{pd}}|\boldsymbol{\mathcal{E}}|^2$,  cannot be neglected, resulting in a total potential

\begin{equation}\label{eq:total U}
\begin{split}
   & U_{\mathrm{tot}}=U_{\mathrm{mag}}+U_{\mathrm{sc}}+U_{\mathrm{pd}} \\
   &=\mu_Bg_{nJ}m_J|\boldsymbol{B}_{\mathrm{eff}}|-\frac{1}{4}(\alpha_{\mathrm{pd}}+\alpha^{\mathrm{sc}}_{nJ}) |\boldsymbol{\mathcal{E}}|^2.
\end{split}
\end{equation}
\noindent Here the ponderomotive and scalar polarisability contributions do not necessarily compensate for each other, because the sign of the scalar polarisability can be either negative or positive.

We obtain the values of the polarisabilities using ARC~\cite{robertson2021arc} and we find that, for $n\geq49$ Rydberg states, the vector polarisability is large enough to create a fictitious magnetic field forming a trap potential of at least 100~$\mu$K depth when the power of the ONF-guided light is $P=10$~mW and the bias magnetic field is $B_{\mathrm{bias}}=30$~G. Guided light powers of a few tens of mW were shown to be feasible experimentally, albeit at different wavelengths. In~\cite{liedl2022observation}, a beam of power 20.5~mW at the wavelength 760~nm was sent through the ONF, while in~\cite{pache2024realization} a beam of power 16.6~mW at the wavelength 685~nm was used. In both experiments, no noticeable damage to the optical nanofibre was observed.

Next, we calculate the total trap potential for a $^{87}$Rb atom in the $\left|49\mathrm{D}_{5/2}, m_J=5/2\right>$ state from Eq.~\ref{eq:total U}, for which $g_{nJ}=1.2$. We use $P= 20$~mW of quasi-circularly polarised 790.2~nm guided light and $B_{\mathrm{bias}}=60$~G applied along the $+y$ direction. Such high values of the power and magnetic field are needed to keep the trap minimum far from the fibre surface due to the fact that the ONF introduces an attractive Casimir-Polder shift for Rydberg states on the order of GHz up to 300~nm away from the fibre~\cite{StourmPRA2020,Vylegzhanin:23}. The calculated trap potential is shown in Fig.~\ref{fig:2d_trap}(a). The minimum of the trap potential is produced approximately 319~nm away from the surface on the left side of the ONF (due to the position of the bias B-field) with the depth of the potential $\sim 207~\mu$K. 

Similarly, we calculate the trap potential for a $^{87}$Rb atom in the $\left|49\mathrm{D}_{5/2}, m_J=5/2\right>$ state created by a QL mode of wavelength 790.2~nm and power  $P=10$~mW with a bias magnetic field $B_{\mathrm{bias}}= 6$~G applied along the $+y$ direction. The bias magnetic field and the power of the guided light are chosen to form the potential at the same distance from the fibre as in the QC case discussed above. The QL mode trap provides  stronger confinement both in radial and  angular directions due to the tightly localised electric field, see Fig.~\ref{fig:2d_trap}(b).  In this configuration, the trap is $\sim0.77$~mK deep and its minimum is formed $\sim 316$~nm away from the ONF surface.

\begin{figure}
    \raggedright{a)}
    
    \centering
    \includegraphics[width=8cm]{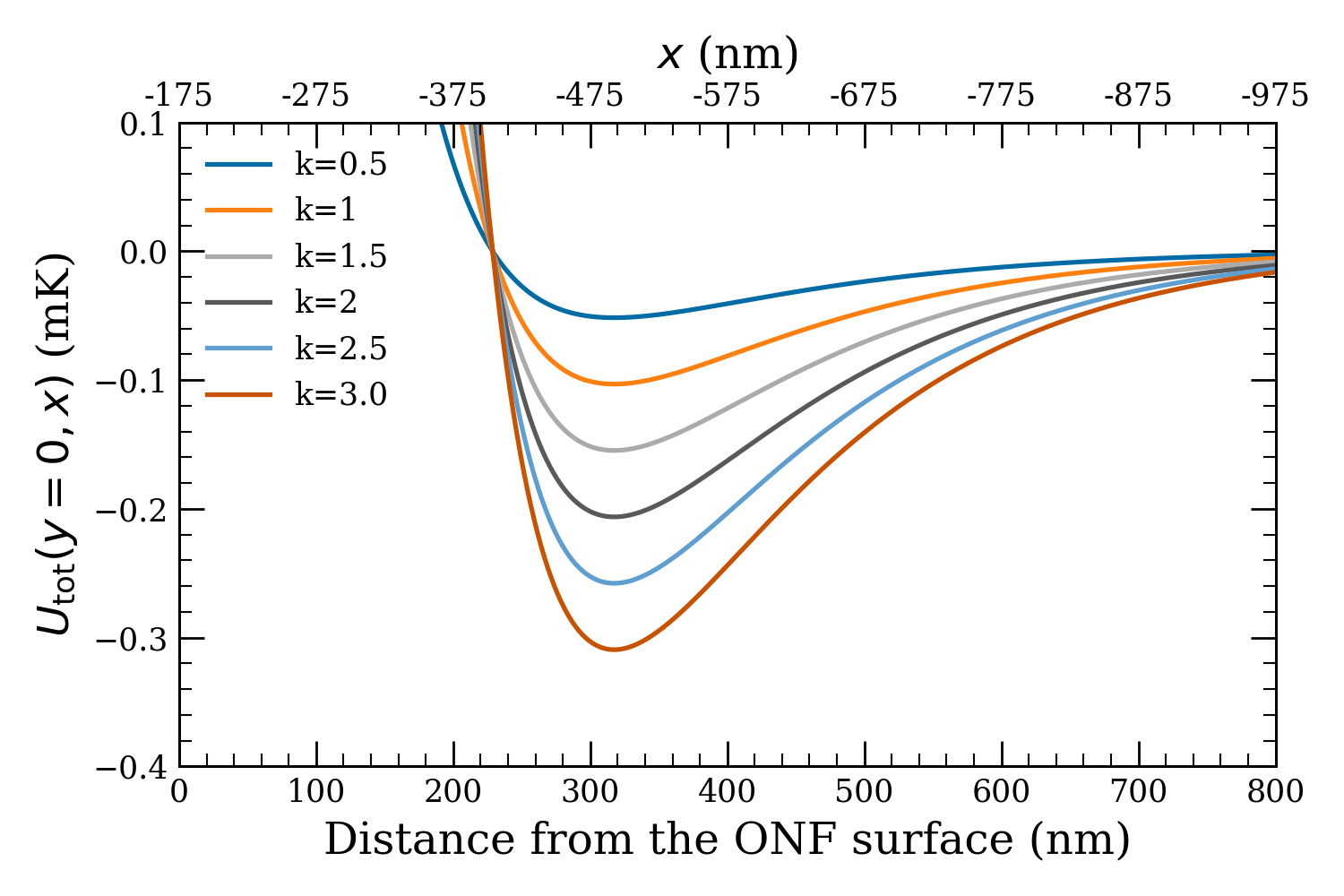}

    \raggedright{b)}

    \centering
    \includegraphics[width=8cm]{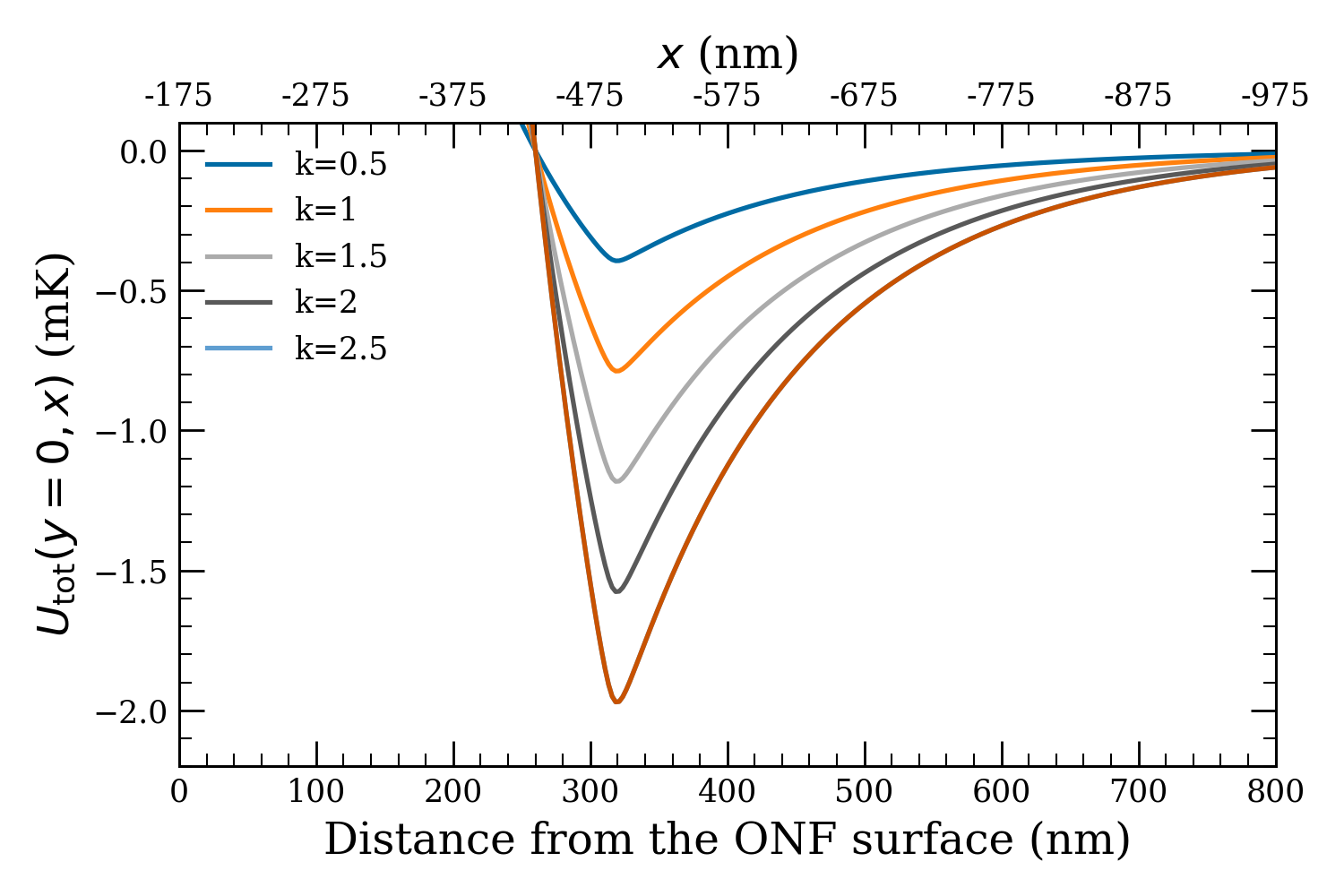}

    \raggedright{c)}

    \centering
    \includegraphics[width=8cm]{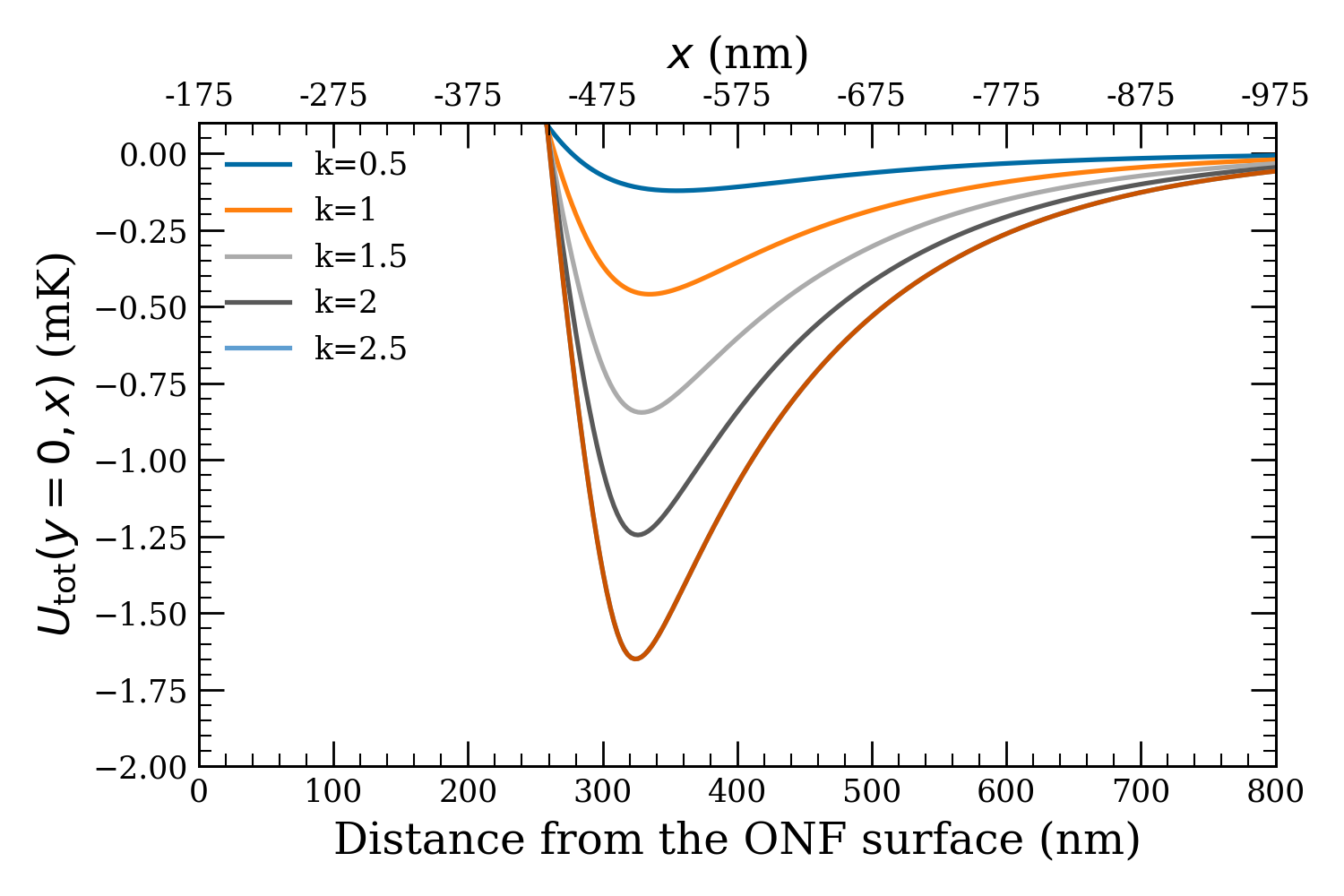}
    \caption{Radial profile of the trap potential, $U_{\mathrm{tot}}(y=0,x)$, for (a) quasi-circularly polarised light with $P=k\times10$~mW, $B_{\mathrm{bias}}=k\times30$~G, and  k=$\{0.5, 1, 1.5, 2, 2.5, 3\}$, and (b,c) quasi-linearly polarised light with power, $P=k\times10$~mW, bias field $B_{\mathrm{bias}}=k\times6$~G, and k=$\{ 0.5, 1, 1.5, 2, 2.5\}$ without and with an additional magnetic field of 2.5~G along the fiber axis, respectively. The fibre radius is $a=175$~nm, the  wavelength is $\lambda= 790.2$~nm, and the Rydberg state is $\left|49\mathrm{D}_{5/2}, m_J=5/2\right>$. The additional $x-$axis provides the positioning related to Fig.~\ref{fig:2d_trap}. The trap is formed on the left side of the ONF as shown in Fig.~\ref{fig:2d_trap}.}
    \label{fig:k_plot}
\end{figure}

The trap depth and minimum position strongly depend on both the strength of the bias magnetic field, $B_{\mathrm{bias}}$, and the power of the ONF-guided light, $P$. These two parameters can be varied experimentally once the Rydberg state is selected. By simultaneously adjusting them so that their ratio is constant, one can change the trap depth while maintaining its minimum position fixed~\cite{schneeweiss2014nanofiber}. To illustrate this, we vary $B_{\mathrm{bias}}$ from 15~G to 90~G and $P$ from 5~mW to 30~mW, maintaining proportionality of the parameters, $P=k\times10$~mW and $B_{\mathrm{bias}}=k\times30$ G. Different trap potentials for  values of $k$ ranging from 0.5 to 3 in steps of 0.5 are shown in Fig.~\ref{fig:k_plot}(a) for a QC mode. Note how the potentials are shifted along the vertical axis so that $U_{\mathrm{tot}} \rightarrow 0$  as $x\rightarrow+\infty$. Each trap minimum is formed 319~nm away from the fibre surface.
The values of  the intensity of the electric field at the trap minimum, $I_0$, the Lamb-Dicke parameter, $\eta=k\sqrt{\hbar/2m\omega}$, where $k=2\pi/\lambda$ and $\omega$ is the smallest trapping frequency, and the radial and azimuthal trap frequencies, $\omega_r$ and $\omega_{\phi}$, respectively are listed  in  Table~\ref{table:params} for all trap configurations considered in Fig.~\ref{fig:k_plot}. An atom can be lost from the trap due to the sign flip of the $m_J$ state when the effective magnetic field reaches nearly zero values. This happens due to the coupling between the motional states of the atom in the trap potential and the Zeeman level energy splitting. In our scheme the spin flip rates, $\Gamma_{\mathrm{sf}}=\frac{\pi \omega_r}{2} \exp\left(-\frac{\mu_B g_{nJ} |\boldsymbol{B}|}{4h\omega_r}\right)$~\cite{sukumar1997spin}, are negligible since the magnetic field is on the order of $\sim 10$~G at the trap minima~\cite{schneeweiss2014nanofiber}.

\begin{table}
\caption{Parameters for the trap configurations shown in Fig.~\ref{fig:k_plot}.  $P$ is the laser power, $B_{\mathrm{bias}}$ the bias magnetic field, $x_0$ the trap distance from the ONF surface, $U_0$ the trap depth, $I_0$ the intensity of the electric field at the trap minimum, Lamb-Dicke parameter, $\eta$, $\omega_r$ and $\omega_{\phi}$ the radial and azimuthal trap frequencies, respectively. All values are as used in Fig.~\ref{fig:k_plot}.}
\label{table:params}
\centering
\begin{tabularx}{\columnwidth}{X X|X X X X X X}
$P$& $B_{\mathrm{bias}}$ & $x_0$ & $U_0$ & $\omega_r/2\pi$ & $\omega_{\phi}/2\pi$ & $~~~\eta$ &$~~I_0$\\ 
 (mw) & (G) & (nm) & ($\mu$K) & (kHz) & (kHz) &  ~~~~~~~~~~(W/mm$^2$) & \\
 \hline
 \multicolumn{8}{>{\hsize=\dimexpr8\hsize+2\tabcolsep+\arrayrulewidth\relax}X}{Quasi-circular polarisation} \\
 \hline
 10 & 30 & 319 & 103 & 120 & 57 & 0.25 & 2629\\  
 15 & 45 & 319 & 155 & 147 & 70 & 0.23 & 3944\\
 20 & 60 & 319 & 207 & 170 & 80 & 0.22 & 5259\\
 \hline
 \multicolumn{8}{>{\hsize=\dimexpr8\hsize+2\tabcolsep+\arrayrulewidth\relax}X}{Quasi-linear polarisation} \\
 \hline
 5 & 3 & 354 & 122 & 153 & 90 & 0.20 & 2601\\
 10 &  6 & 334 & 460 & 367 & 193 & 0.14 & 6021\\
 15 & 9 & 329 & 846 & 569 & 287 & 0.11 & 9372
\end{tabularx}

\end{table}
Likewise, for the QL mode, we scale the power and bias field such that $P=k\times10$~mW and $B_{\mathrm{bias}}=k\times6$~G for $k=\{ 0.5, 1, 1.5, 2, 2.5\}$, see Fig.\ref{fig:k_plot}(b). The sharp potential observable in Fig.\ref{fig:k_plot}(b) arises from the fictitious field being almost fully cancelled by the constant bias field due to the fact that the fictitious field is only generated in the $xy-$plane, see eq.~\ref{eq:QL_B}. In contrast, for the QC polarised light, there is a light-induced fictitious magnetic field component along the $z-$axis, therefore  $B_{\mathrm{eff}}$ never reaches zero and  $U_{\mathrm{tot}}$ has a smooth shape [see Fig.~\ref{fig:k_plot}(a)]. In addition, the fact that the total magnetic field reaches almost zero at the local minimum for quasi-linearly polarised light could lead to high rates of spin flipping~\cite{schneeweiss2014nanofiber}, i.e., changes to the sign of $m_F$ state, which would be detrimental to the trap lifetime - once an atom changes the sign of its $m_F$ state it is repelled from the trap. For example, for $P=5$~mW and $B_{\mathrm{bias}}=3$~G the spin flip rate, $\Gamma_{\mathrm{sf}}$, is on the order of $10^7$~s$^{-1}$. To avoid the sharp feature of the potential and high spin flip rates we add an additional magnetic field, $B_{\mathrm{add}}\approx 2.5$~G, along the fiber axis, see Fig.\ref{fig:k_plot}(c). It decreases dramatically the spin flip rate to the order of $10^{-2}$~s$^{-1}$, however the trap depth is decreased as well. Each trap minimum is  formed around 340~nm away from the fibre surface.  We compute $I_0$, $\omega_r$, $\omega_{\phi}$, and list them in  Table~\ref{table:params}. The additional magnetic field also has the advantage of providing a defined quantisation axis in the trap, allowing for the possibility to study state and orientation specific phenomena such as the possible restrictions on angular momentum states next to an ONF.  

Note that for the trapping parameters presented in Table~\ref{table:params},  the rate of Raman scattering, $R_\mathrm{sc}$, due to the far-off detuned trapping field is of the order of $10^{-11}$ s$^{-1}$ for the considered Rydberg state. Therefore, we assume that the lifetime of the Rydberg state in the trap is limited by the black body radiation and is $\sim100~\mu$s.  The values of $\eta$  show that we are in the upper limit of the Lamb-Dicke regime, therefore there is a low though nonvanishing probability of motion excitation.

Hereafter, we focus only on the QC  mode case, since it can be approximated by the harmonic oscillator near the minimum of the trap. In the QL mode case, the anharmonicity is much stronger, which can lead to an increase in escape rates. However, analysis of the anharmonicity is not the goal of this study.

\section{Towards trapping ground and Rydberg states}\label{magic trap}
One of the main requirements for a quantum technology platform based on Rydberg atoms is the ability to keep an atom trapped both in the ground and Rydberg state during a quantum gate operation~\cite{zhang2011magic}, which usually takes a few $\mu$s~\cite{su2016one}. So far, there are a limited number of approaches to do so, since trapping each state generally requires a different trapping scheme~\cite{zhang2011magic, cortinas2020laser}. In this section, we investigate the possibility of creating a comparable trap potential for both Rydberg and ground state $^{87}\mathrm{Rb}$ atoms using the same fictitious magnetic field trap. This requires the total trap potential for a ground state atom, $U_\mathrm{G}$, to be as close as possible to that for a Rydberg state atom, $U_\mathrm{R}$, i.e.,
\begin{equation}\label{eq:9}
\begin{split}
&\mu_Bg_{nJ}m_J|\boldsymbol{B}_{\mathrm{eff,R}}|+U_{\mathrm{sc, R}}+U_{\mathrm{pd}}\\ &\approx\mu_Bg_{nJF}m_F|\boldsymbol{B}_{\mathrm{eff,G}}|+U_{\mathrm{sc,G}},
\end{split}
\end{equation}
\noindent where $\mathrm{G}$ and $\mathrm{R}$ denote the ground and Rydberg states, respectively.
Here, we choose the QC-polarised guided mode, since it leads to a trap potential which is closer to harmonic than its QL counterpart. Moreover, the QL mode  creates a tighter trap, which may not be ideal for higher Rydberg states in alkali atoms due to the effective atom size, which scales as $n^2$.

As the bias field is fixed, the fictitious fields for both the ground and Rydberg states must have the same sign and order of magnitude to produce similar effective potentials. From eqs.~(\ref{eq:1},\ref{eq:2}) one gets:

\begin{equation}\label{eq:13}
    \frac{\alpha^{\mathrm{v}}_{nJF}}{ g_{nJF} F} \approx \frac{\alpha^{\mathrm{v}}_{nJ}}{ g_{nJ} J},
\end{equation}
where $\alpha^{\mathrm{v}}_{nJ}$ is the vector polarisability of a chosen Rydberg state and $\alpha^{\mathrm{v}}_{nJF}$ is the vector polarisability of the ground state. The latter can be calculated from \cite{le2013dynamical}
\begin{equation}\label{eq:13}
\begin{split}   
    &\alpha^{\mathrm{v}}_{nJF}=(-1)^{J+I+F+1}\sqrt{\frac{2F(2F+1)(J+1)(2J+1)}{2J(F+1)}} \\
    &\times\begin{Bmatrix}
    F&1&F\\
    J&I&J
    \end{Bmatrix}\alpha^{\mathrm{v}}_{nJ},
\end{split}    
\end{equation}
\noindent where $\alpha^{\mathrm{v}}_{nJ}$ can be calculated from ARC~\cite{robertson2021arc}.

We set the ground and Rydberg states to be $\left|\mathrm{5S}_{1/2}, F=1 \right>$ and  $\left|n\mathrm{G}_{9/2} \right>$,  respectively, with $g_{nJF}=-0.5$, $\alpha^{\mathrm{v}}_{nJF}=62\times10^{-5}$  and  $g_J=10/9,~ n>20$. The $n\mathrm{S}$ and $n\mathrm{P}$ Rydberg states never attain vector polarisability values, $\alpha^{\mathrm{v}}_{nJ}$, comparable to that for the ground state for the chosen wavelength; therefore,  it is impossible to produce a fictitious magnetic field that is similar for the ground state and the $n$S or $n$P  Rydberg states. 

We used the ARC~\cite{robertson2021arc} for $n\in[20,80]$ to find a Rydberg $\left|n\mathrm{G}_{9/2}\right>$ state for which (Eq.~\ref{eq:13}) holds true at $\lambda=790.2~$nm. We set the limit to $n=80$ so that the Rydberg atom electron wave function, with an approximate mean distance from the atomic core of 450~nm, does not overlap with the ONF. We found that the first Rydberg state to fulfil the condition from Eq.~\ref{eq:13} is $\left|\mathrm{68G}_{9/2} \right>$ for which $\alpha^{\mathrm{v}}_{nJ}=32\times10^{-5}$. We used the $F=1$ hyperfine ground state so that the fictitious magnetic fields for both the Rydberg and the ground states have the same direction, hence the trap is produced on the same side of the nanofibre. Higher values of $\alpha^{\mathrm{v}}_{nJ}$ may be achieved for Rydberg states of higher principal quantum number, $n$. However, the effective size of a Rydberg atom scales as $n^2$; hence, the larger the effective atom size, the further away from the ONF it must be trapped to avoid additional energy shifts from atom-surface interactions\cite{kubler2010coherent, StourmPRA2020,Vylegzhanin:23,wongcharoenbhorn2023casimir}. Therefore, we limit our discussion on the trap for both ground and Rydberg states to the $\left|\mathrm{68G}_{9/2}\right>$ state, noting that excitation to the $\left|\mathrm{68D}_{5/2}\right>$ state from the nanofibre-guided field has already been observed experimentally~\cite {Vylegzhanin:23} and this state can be microwave frequency coupled to $\left|\mathrm{68G}_{9/2}\right>$~\cite{signoles2014confined}.

To compensate for any differences in the magnetic potentials, one can detune the trap wavelength from the tune-out condition, $\lambda_{to}=790.2~$nm, to finely adjust the scalar potential of the ground state. The magnetic part of the ground state trap potential does not change because the vector polarisability is changed by less than 1$\%$. The Rydberg state potential is also unmodified since, for a fine adjustment of the far-off resonant wavelength, the vector polarisability is negligibly affected.

\begin{figure}[h!]
    \centering
    \includegraphics[width=8.5cm]{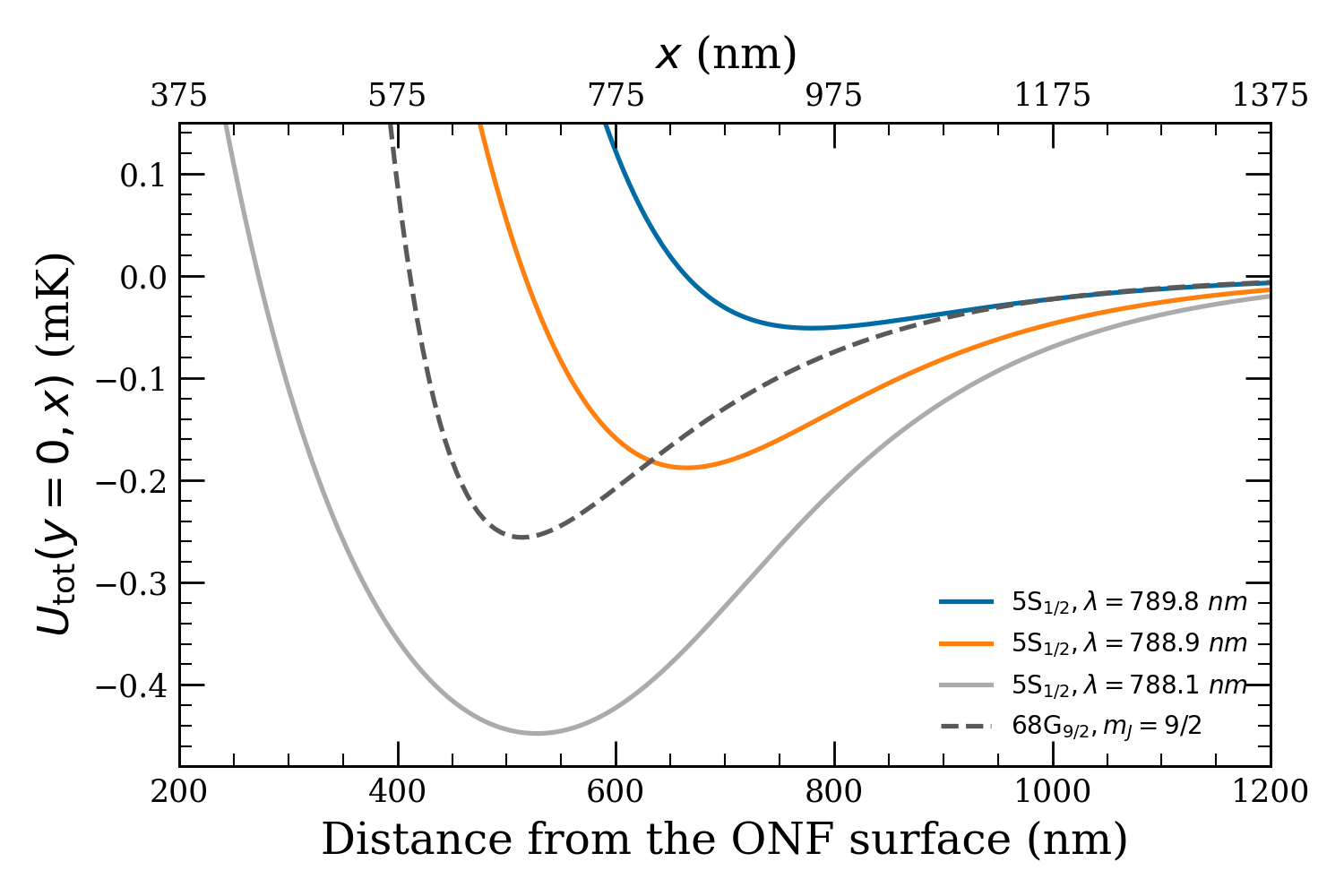}

    \caption{Radial profile of the trap potential, $U_{\mathrm{tot}}(y=0,x)$, for the $\left|\mathrm{68G}_{9/2}, m_J=9/2\right>$ Rydberg state (black dashed line) and  the $\left|\mathrm{5S}_{1/2}, F=1, m_F=-1\right>$ ground state for quasi-circularly polarised light with a  wavelength of 789.8~nm (blue line), 788.9~nm (orange line), and 788.1~nm (gray line). $P =15$~mW, $B_{\mathrm{bias}} = 15$~G, and the fibre radius is $a =175$~nm.}
    \label{fig:magic_trap}
\end{figure}

To minimise the difference between trap potentials for ground and Rydberg states, we vary the three free parameters, i.e.,  the power, $P$, and the detuning, $\Delta_{790}$, of the 790.2~nm light, and the amplitude of the bias magnetic field, $B_{\mathrm{bias}}$. We note that both $P$ and $B_{\mathrm{bias}}$ shift the two  trap potentials in the same direction and, therefore, cannot be used to improve the overlap between them. The one parameter which allows us to shift the ground state potential while keeping the Rydberg state potential fixed is $\Delta_{790}$. We calculate the potentials for both states with  $P=15$~mW and $B_{\mathrm{bias}}=15$~G and then vary $\Delta_{790}$ (see Fig.~\ref{fig:magic_trap}). For a wavelength $\lambda = 788.1$~nm, the trap potential minima for the Rydberg and the ground state are less than 10~nm apart and the difference in their trap depths is around 0.18~mK. The ratio of the trap depths for the two states is around $40\%$. We estimated the lifetime of the ground state atoms in the trap to be $\sim20$~s from $\tau_{\mathrm{trap}}=U_0/(2E_{\mathrm{rec}}R_{\mathrm{sc}})$~\cite{le2021optical}. Here, $U_0$ is the depth of the trap potential, and $E_{\mathrm{rec}}=(\hbar k)^2/2M$ is the recoil energy of a photon scattered from an atom of mass $M$. The upper limit of the atom lifetime in the trap is set by recoil heating~\cite{goban2012demonstration}, however, Raman scattering can contribute to an additional loss. Due to off-resonant Raman scattering, an atom may undergo a change in the $m_F$ state that affects the trap potential. Raman scattering gives the worst case estimate of the atomic lifetime in the trap, $\tau_\mathrm{R}\sim1/R_\mathrm{sc}$. The coherence time is also limited by scattering processes, the dominant one being Raman scattering, since the trap wavelength lies between the $^{87}$Rb D1 and D2 lines. In our configuration, typical values of $R_\mathrm{sc}$ are on the order of $50$~s$^{-1}$, setting the worst-case trap lifetime limit to $\sim20$~ms. This trapping time is sufficient for many Rydberg experiments~\cite{lee2019coherent}. For the Rydberg state, the scattering rate is on the order of $10^{-11}~$s$^{-1}$ and, therefore, the guided light does not cause sufficient scattering to heat the atoms during a typical experiment duration. Thus, the trap lifetime of Rydberg states is primarily limited by the black body radiation and is on the order of 100~$\mu$s~\cite{gallagher2005rydberg,robertson2021arc}.

\begin{figure}[h!]
    \centering
    \includegraphics[width=8.5cm]{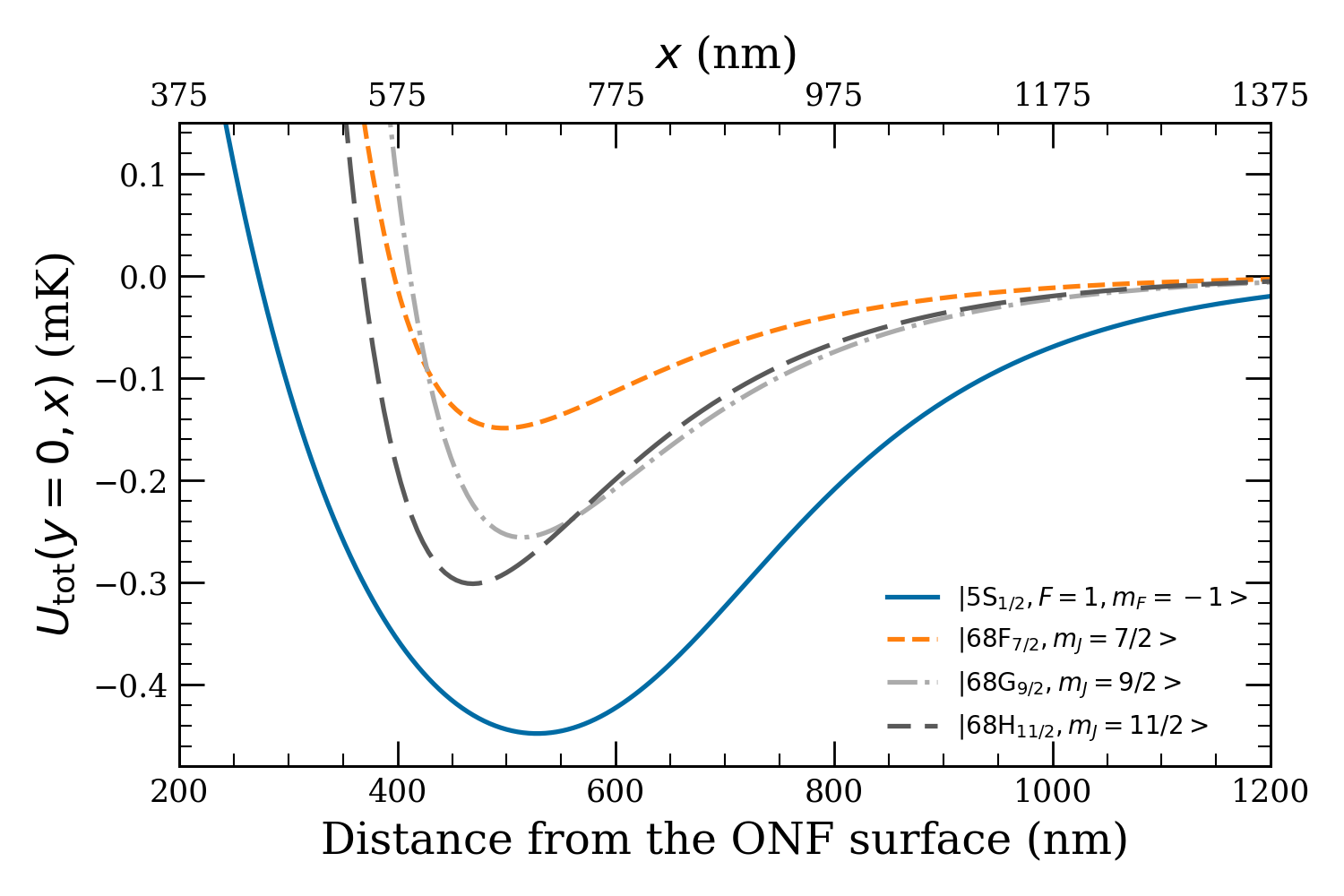}

    \caption{Radial profile of the trap potential, $U_{\mathrm{tot}}(y=0,x)$, for the $\left|\mathrm{68F}_{7/2}, m_J=7/2\right>$ (orange dashed line), $\left|\mathrm{68G}_{9/2}, m_J=9/2\right>$ (grey dot dashed line), $\left|\mathrm{68H}_{11/2},  m_J=11/2\right>$ (black dashed line) Rydberg states, and for the $\left|\mathrm{5S}_{1/2}, F=1, m_F=-1\right>$ ground state (solid blue line) for quasi-circularly polarised light.  $P=15$~mW, $B_{\mathrm{bias}}=15$~G, the fibre radius is $a = 175$~nm, and the wavelength is $\lambda=788.1$~nm. }
    \label{fig:more circular}
\end{figure}

To reduce the difference between the depths of the traps  associated with the two states, one can exploit other angular momentum quantum number Rydberg states, such as $n$F, $n$H, etc., since they can have different vector polarisabilities and higher $m_J$ quantum numbers.  Therefore, they may experience a deeper magnetic potential. We show the total trap potentials for 68F, 68G, and 68H Rydberg states in comparison to the $\left|\mathrm{5S}_{1/2}, F=1, m_F=-1\right>$ ground state trap potential for $P=15$~mW, $B_\mathrm{bias}=15$~G, and $\lambda=788.1$~nm in Fig.~\ref{fig:more circular}. The spacing between the positions of the minima of the trap potentials for the ground state and the $\left|\mathrm{68F}_{7/2}, m_J=7/2\right>$ ($\left|\mathrm{68H}_{11/2}, m_J=11/2\right>$) Rydberg state is around 10~nm (40~nm) and the difference in the trap depths is around 0.31~mK (0.16~mK).

\begin{figure}[h]
    \raggedright {a)}

    \centering
    \includegraphics[width=8.3cm]{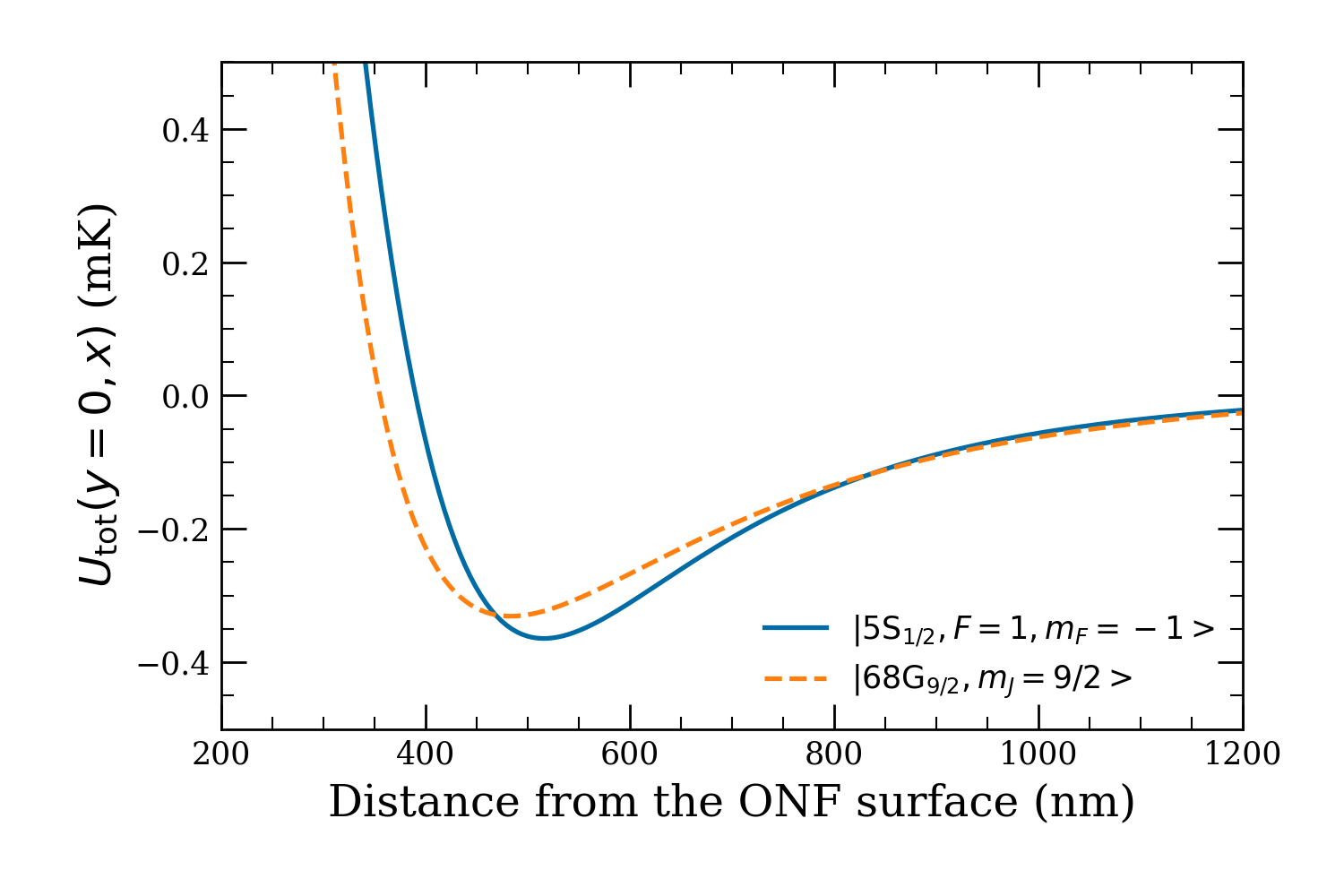}

    \raggedright{b)}
    
    \centering
     \includegraphics[width=8cm]{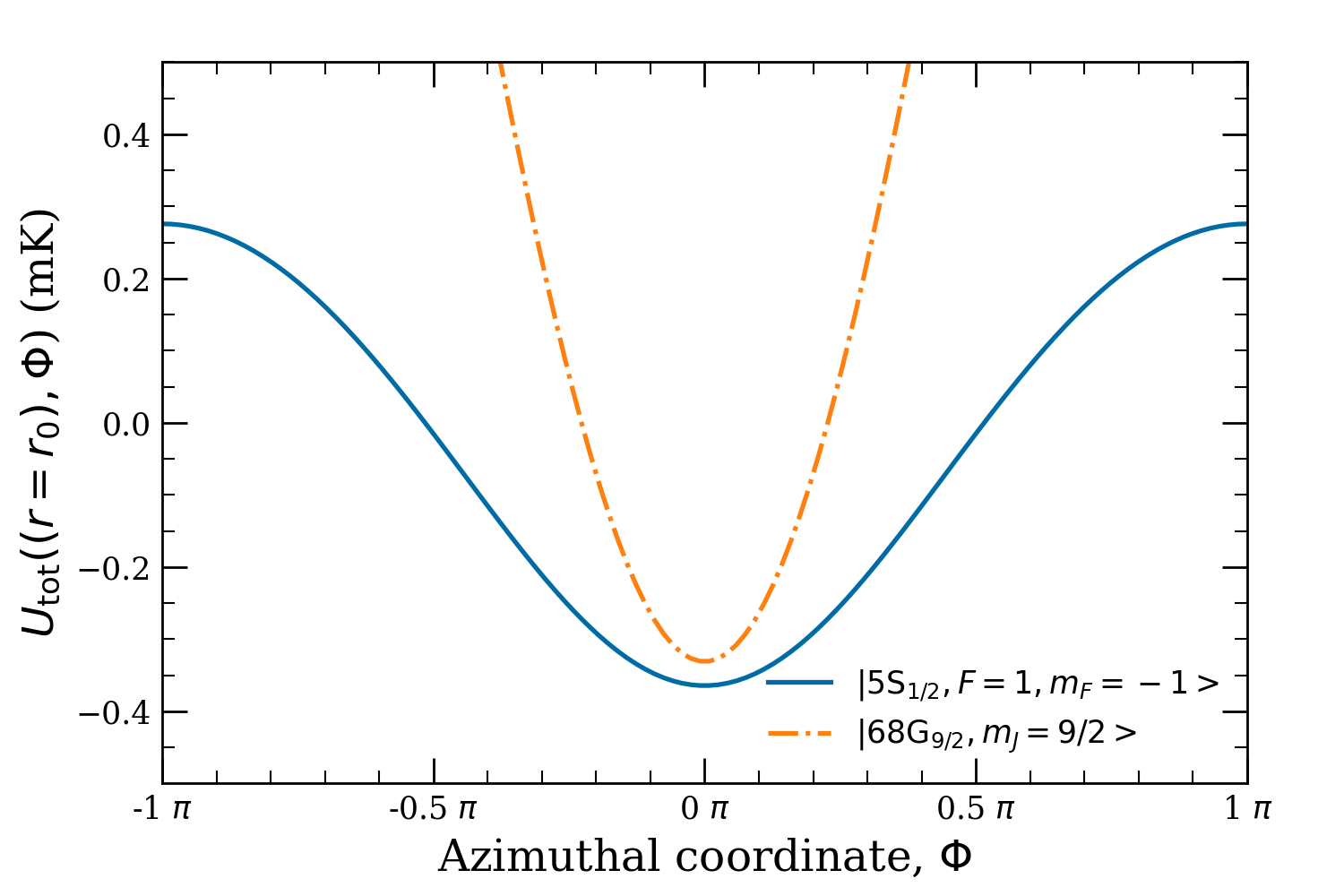}

    \caption{(a) Radial and (b) azimuthal profiles of the total trap potentials, $U_{\mathrm{tot}}$, for the $\left|\mathrm{68G}_{9/2}, m_J=9/2\right>$ Rydberg state (orange dashed line) and  the $\left|\mathrm{5S}_{1/2}, F=1, m_F=-1\right>$ ground state (blue line) for $P_1=12$~mW of quasi-circularly polarised light with $\lambda_1=789.7$~nm and $P_2=6.3$~mW of quasi-circularly polarised light with $\lambda_2=1015$~nm. $B_{\mathrm{bias}}=45$~G and the fibre radius is $a=175$~nm.}
    \label{fig:state insensitive}
\end{figure}

Another approach to reduce the differences in the trap potentials is to guide light of a second  wavelength, $\lambda_2$, through the ONF, in addition to the original wavelength, which henceforth we refer to as  $\lambda_1$. If $\lambda_2=1015$~nm, an attractive scalar potential for the ground state is produced, with an almost zero fictitious magnetic field. For the Rydberg state, $\lambda_2$ produces a small repulsive scalar potential and a nonzero fictitious magnetic field. We run an optimisation algorithm via the Python {SciPy} library~\cite{virtanen2020scipy} to minimise the trap  depth differences introduced by
\begin{equation}\label{eq:optimiz}
\begin{split}
    &\Delta U_0=\mu_Bg_{nJ}m_J|\boldsymbol{B}_{\mathrm{eff, R}}|_+U_{\mathrm{sc}(\lambda_1),\mathrm{R}}+U_{\mathrm{sc}(\lambda_2),\mathrm{R}} \\
    &+U_{\mathrm{pd}(\lambda_1)}+U_{\mathrm{pd}(\lambda_2)} \\
    &-\mu_Bg_{nJF}m_F|\boldsymbol{B}_{\mathrm{eff,G}}|+U_{\mathrm{sc}(\lambda_1),\mathrm{G}}+U_{\mathrm{sc}(\lambda_2),\mathrm{G}},
\end{split}
\end{equation}
\noindent where the free parameters are the detuning of $\lambda_1$, guided light powers of $\lambda_1$ and $\lambda_2$, and the strength of the bias magnetic field. Here, $\lambda_1$ and $\lambda_2$ are the wavelengths of light in vacuum used to produce both the scalar AC Stark shift potential, the effective magnetic field potential, and the ponderomotive potential. We tuned the value of $\lambda_1$ from 788.1~nm to 789.7~nm and calculated a trap potential with $B_{\mathrm{bias}}=45$~G, $P_{\lambda_2}=6.3$~mW, and $P_{\lambda_1}=12$~mW, see Fig.~\ref{fig:state insensitive}.

The ratio of the trap depths for the ground and the Rydberg states in the trap is around $10\%$ and the difference in the position of the trap minima is less than 30~nm. The radial trap frequencies are 146~
kHz and 174~kHz and the azimuthal trapping frequencies are 62~kHz and 32~kHz for the Rydberg state and the ground state, respectively. The scattering rate, $R_{\mathrm{sc}}$, for the Rydberg state is on the order of $10^{-16}~\mathrm{s}^{-1}$ and $10^{-11}~\mathrm{s}^{-1}$ for the 1015~nm and 789.7~nm light, respectively, since  both wavelengths are far detuned from the nearest auxiliary transition $\left|68\mathrm{G}_{9/2}\right>\rightarrow\left|4\mathrm{F}_{7/2}\right>$; therefore, the heating of the Rydberg state is negligible. The lifetime of ground state atoms in the trap is around 20~ms. The lifetime of the Rydberg state atoms in the trap is primarily limited by the black body radiation and is on the order of 100~$\mu$s~\cite{gallagher2005rydberg,robertson2021arc}. The spin flip rate for  the Rydberg state is negligibly small. The parameters of the trap for both states are shown in  Table~\ref{tab:params_ground_and_rydb}.

\begin{table}
\caption{Parameters for the trap configurations shown in Fig.~\ref{fig:k_plot}. $x_0$ is the trap distance from the ONF surface, $U_0$ is the trap depth, $\omega_r$ and $\omega_{\phi}$ are the radial and azimuthal trap frequencies, respectively, $I_{\lambda_1}$ and $I_{\lambda_1}$ are the intensities of the electric field at the trap minimum for 789.7~nm and 1015~nm, respectively, $\eta$ is the highest value Lamb-Dicke parameter among two trapping wavelengths.} All values are as used in Fig.~\ref{fig:state insensitive}.
\label{tab:params_ground_and_rydb}
\centering
\begin{tabularx}{\columnwidth}{X |X X X X X X X X}
 & $x_0$ & $U_0$  & $\omega_r/2\pi$  & $\omega_{\phi}/2\pi$ & $   I_{\lambda_1}$ & $I_{\lambda_2}$ & $~\eta$\\ 
 & (nm) & ($\mu$K) &  (kHz) & (kHz) & $~~$(W/mm$^2$) &   \\

 \hline
 $5$S$_{1/2}$ & 515 & 364 & 174 & 32 & 854 & 1009 & 0.34 \\  
 $68$G$_{9/2}$ & 484 & 330 & 146 & 62 & 1040 & 1105 & 0.25 
\end{tabularx}
\end{table}

The potentials described above are translationally invariant along the nanofibre, creating a guiding structure for atoms parallel to the fibre axis. Axial confinement may be introduced using an externally applied inhomogeneous magnetic field. Furthermore, counterpropagating fibre-guided fields can generate a periodic modulation of the fictitious magnetic field~\cite{le2013state}, enabling the formation of a periodic array of trapping sites.

In principle, one can find other Rydberg states for which the two-colour approach allows for trapping both ground and Rydberg state atoms.

\section{Quadrupole AC Stark shift and impact of the finite size of Rydberg atoms}
\begin{figure*}
    \centering
    \includegraphics[width=17cm]{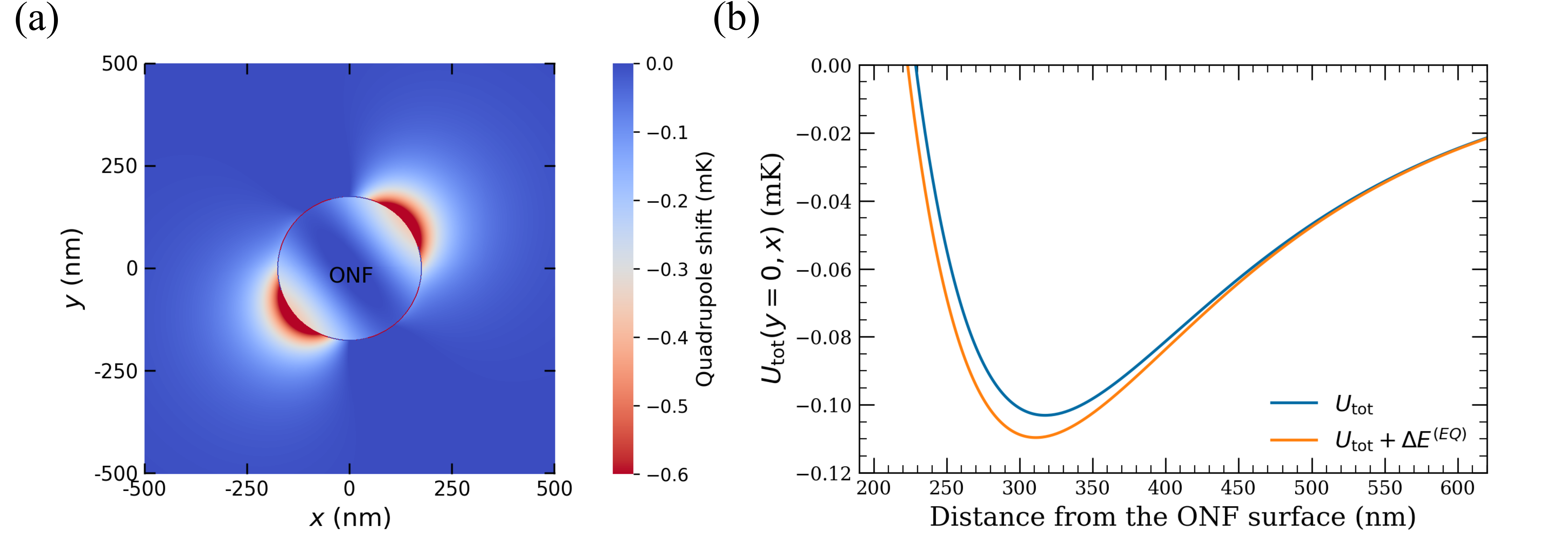}
    \caption{(a) 2D plot of the quadrupole AC Stark shift for  10~mW of  790.2~nm quasi-circularly polarised light. (b) 1D plot of the trapping potential, $U_{\mathrm{tot}}(y=0,x)$, for the $\mathrm{49D}_{5/2}$ Rydberg state for 10~mW of  790.2~nm quasi-circularly polarised light and $B_{\mathrm{bias}}=30$~G: blue line - without the quadrupole shift, orange line -  with the quadrupole shift. The fiber radius is $a=175$~nm.}
    \label{fig:impact}
\end{figure*}

Rydberg atoms are unique compared to lower state atoms since their valence electron wave function can extend to significant distances from the nucleus, no longer allowing us to treat the atom as a point particle~\cite{zhang2011magic}. When we consider the significant gradient of the evanescent field of the ONF, it is clear that the electric field intensity varies dramatically across the valence electron wave function. Additionally, as proposed and demonstrated in some earlier works~\cite{quadrupole_excitations, le2022transfer, le2023direction}, the gradient of the evanescent field of an optical nanofibre is sufficient to excite quadrupole transitions even at relatively low laser powers.  
In this section, we calculate the quadrupole energy shift due to the evanescent field of the ONF, and include the effects of weighting the potential over the electron wave function similar to that in ~\cite{zhang2011magic}.

We follow the derivation in~\cite{PhysRevA.98.013406} for the quadrupole light shift, but we include all components of the tensor product $\{\{\nabla \mathbf{E}^*\}_2 \otimes \{\nabla \mathbf{E}\}_2\}_{k,0}$ due to the complex nature of the polarisation of the evanescent field. The expression for the energy shift is

\begin{equation}
\begin{split}
     &\Delta E_{nFJIM}^{\mathrm{(EQ)}}=\frac{1}{4}\sum_{k=0}^{4}(-1)^k \{\{\nabla \mathbf{E}^*\}_2 \otimes \{\nabla \mathbf{E}\}_2\}_{k,0}\\ &\sqrt{2k+1}(-1)^{F-M}
     \begin{pmatrix}
         F&F&k\\
         -M&M&0
     \end{pmatrix}
     \sum_{n',F',J'}
     \begin{Bmatrix}
         F&F&k\\
         2&2&F^\prime
     \end{Bmatrix}
     \times\\
     &|Q^{(2)}_{nFJI,n'F'J'I'}|^2R^{(k)}_{n,F,J;n',F',J'}
\end{split}\label{eq:quadrupole}
\end{equation}
where $Q^{(2)}_{nFJI,n'F'J'I'}$ is the reduced matrix element of the quadrupole operator, \textbf{Q}, between two states defined by the set of quantum numbers $n,F,J,I,M$ and 
\begin{equation}
\begin{split}
    &R^{(k)}_{n,F,J;n',F',J'}=\frac{1}{\hbar(\omega_{nFJ}-\omega_{n'F'J'}+\omega)}+\\
    &(-1)^k\frac{1}{\hbar(\omega_{nFJ}-\omega_{n'F'J'}-\omega)}.
\end{split}
\end{equation}

We numerically calculate the gradients of the electric field of a QC polarised mode, from which we deduce the energy shift via Eq.~(\ref{eq:quadrupole}). The calculated light shift in mK for the $\mathrm{49D_{5/2}}$ state for 10~mW of 790.2~nm light power is shown in Fig.~\ref{fig:impact}(a). The quadrupole light shift can modify the potential on the order of 0.5~mK at distances of approximately 100~nm from the ONF surface; however, it is negligible at distances larger than 400~nm from the ONF surface [see Fig.~\ref{fig:impact}(b)].


An additional complication of highly excited Rydberg states in alkali atoms located near optical nanofibres is that their effective size, i.e., the distance of their valence electron from the nucleus, becomes comparable to the atom-fibre separation. Since the evanescent field amplitude changes quite rapidly from the surface of the fibre, it varies significantly across the atom and, therefore,  a point dipole approximation may not be valid. To attempt to take into account the size of the Rydberg atom, we demonstrate how the ponderomotive potential may be modified due to the variation of the evanescent field across the atom. By following the procedure in~\cite{zhang2011magic, topcu2013intensity} we can write

\begin{equation}\label{eq:weighting}
\begin{split}
    &U_{\mathrm{pd_\psi}}(\textbf{R})=\int d^3r U_{\mathrm{pd}}(\textbf{R}+\textbf{r})|\psi(\textbf{r};\textbf{R})|^2 \\
    & =\frac{e^2}{2\epsilon_0 c m_e \omega^2}\int d^3r I(\textbf{R}+\textbf{r}) |\psi(\textbf{r};\textbf{R})|^2
\end{split}
\end{equation}
where \textbf{R} is the atom nuclear coordinate, \textbf{r} is the coordinate of the electron relative to the nucleus, $\mathrm{\psi(\textbf{r})}$ is the radial part of the electron wave function, and $U_{\mathrm{pd}}$ is the ponderomotive potential when treating the atom as a point dipole.

\begin{figure}
    \centering
    \includegraphics[width=8cm]{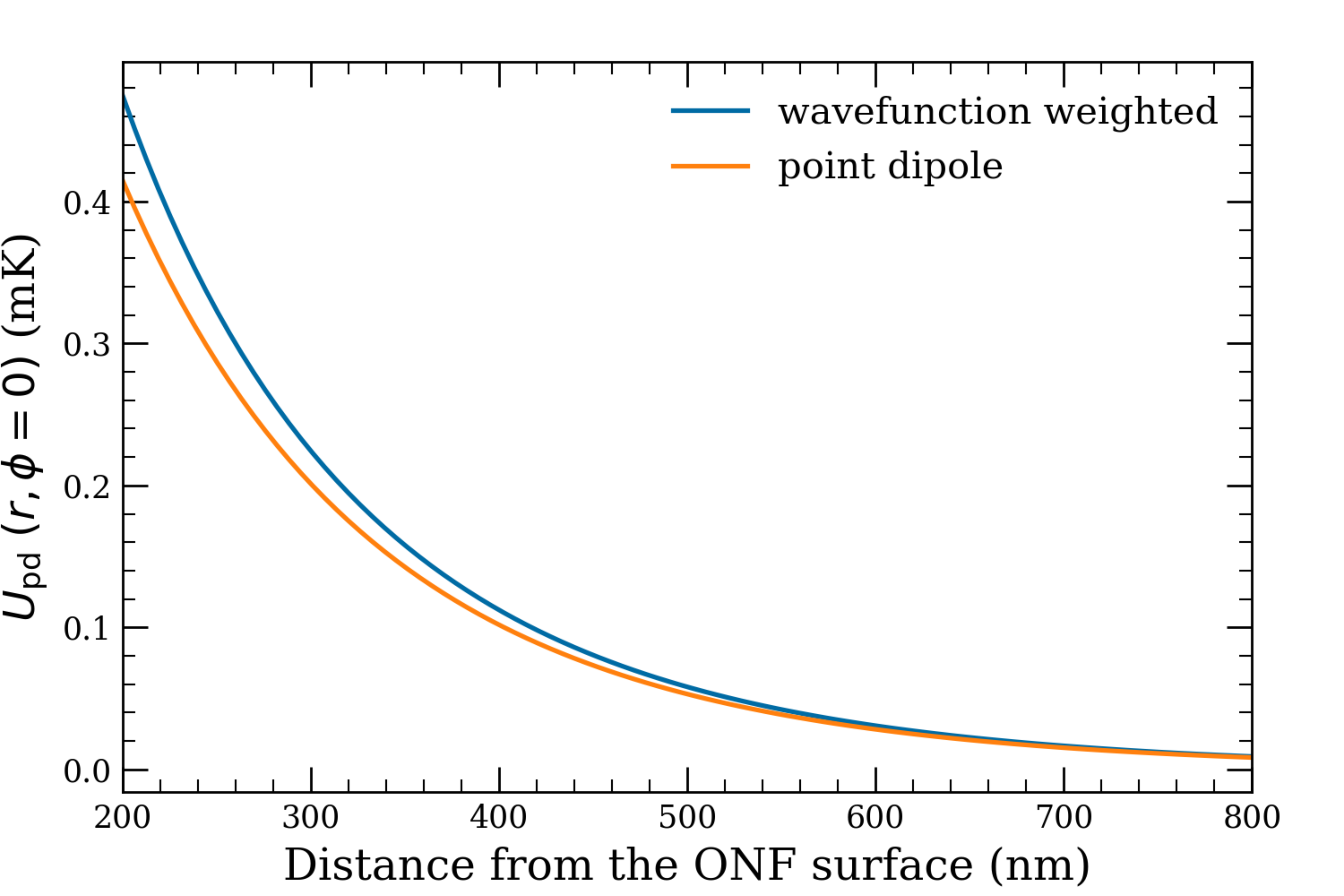}
    \caption{Wave function adjusted (blue line), $U_{\mathrm{pd_\psi}}$, and point dipole (orange line), $U_{\mathrm{pd}}$, calculation of the ponderomotive potential for the $\mathrm{49D_{5/2}}$ Rydberg state for 10~mW of 790.2~nm quasi-circularly polarised light. The fibre radius is $a=175$~nm.}
    \label{fig:ponder}
\end{figure}

We calculate the wave function adjusted potential for the $\mathrm{49D}_{5/2}$ state.  We consider only the radial part of the electron wave function and average out the angular part due to the unknown distribution of the wave function relative to the $z-$axis, thereby simplifying the calculations.  In Fig.~\ref{fig:ponder}, we plot the ponderomotive potential for the QC mode for a propagating laser power of 10~mW. When an atom is close to the ONF surface, where the gradient of the electromagnetic field is steep, the size of the atom has a reasonable impact on the potential experienced. However, at distances greater than approximately 450~nm from the fibre surface, the ponderomotive potential tends towards that calculated treating the atom as a point dipole.  For Rydberg states with $\ell\geq 1$ the situation might be more complex than what we have assumed, as the electron wave function can no longer be treated as spherically symmetric.  The projection of the angular momentum, $J$, and the orientation of the atom relative to the ONF, may play a role in the wave function integration of Eq.~\ref{eq:weighting}. However, the quantification of the angular wave function distribution due to the potential effects of the ONF presence is beyond the scope of this work.

The shape and strength of the light-induced fictitious magnetic field, however, are independent of the size of a Rydberg atom. This is due to the nature of the AC Stark shift, which is primarily determined by the overlap between the wave functions of the ground state and the Rydberg state’s valence electron. This overlap is significant only near the position of the ground state atom, as the ground state's electron wave function rapidly decays just a few nanometres from its peak value~\cite{foot2005atomic}.

\section{Conclusions}
In conclusion, we have calculated trap potentials for $^{87}$Rb atoms in high-level Rydberg states using the fictitious magnetic field generated from the evanescent field of an optical nanofibre. We presented important parameters such as trap depths, trap minimum positions, and trapping frequencies for various powers of the trapping light and strengths of the bias magnetic field for a given Rydberg state. We analysed and compared the trap configurations for both quasi-linearly and quasi-circularly polarised guided light in the ONF. We concluded that the QL guided mode configuration creates a deeper trapping potential due to the spatial profile of the mode; however, it requires an additional bias magnetic field, transverse to the bias field, to decrease the rate of spin flips that would lead to atom loss from the trap.

Additionally, we described a configuration allowing one to effectively trap both Rydberg and ground state $^{87}$Rb atoms in comparable traps. This could enable the study of  fundamental properties of Rydberg atoms, such as  Rydberg blockade, the Rydberg state lifetime, the Casimir-Polder interaction at well-defined distances from a dielectric surface~\cite{Stourm_2019,StourmPRA2020,Stourm_2023}, and could enable us to study 1D chains of trapped Rydberg atoms~\cite{brion2020floquet}. The ability to have both states trapped during  various Rydberg experiments could decrease spatial dephasing~\cite{saffman2011rydberg} during the operation time, increasing the fidelity and reproducibility of these measurements. We showed that higher angular momentum Rydberg states could be used to reduce the differences in the trapping depths of the ground and Rydberg state potentials. Such Rydberg states can be reached via microwave transitions from the $n\mathrm{D}$ Rydberg state~\cite{bai2024microwave}.  By adding a second guided light field at a different wavelength, further tuning of the two trapping potentials is viable, leading to a better overlap in the minima positions and depths, and decreasing the off-resonant scattering of the ground state. Finally, we calculated the quadrupole AC Stark shift of the Rydberg levels near the ONF as well as the effect of the size of the Rydberg atom on the shape of the ponderomotive potential when the electromagnetic field varies significantly across the atom.

To load a magnetic trap as discussed here, one could transfer ground state atoms from a two-colour ONF trap by adiabatically transforming from one potential to another~\cite{schneeweiss2014nanofiber}. In addition, one could adiabatically load atoms into the QL mode trap configuration directly from a magneto-optical trap (MOT) as  done for conventional wire traps~\cite{fortagh1998miniaturized}. The usual temperature of $^{87}$Rb atoms in a MOT is around 140~$\mu$K, whereas the depth of the trap can reach 1.5~mK with an experimentally reasonable magnetic field.

Our proposed trapping scheme can be used in experiments not requiring Zeeman splitting and trapping of different $m_F$ and $m_J$ levels in similar potentials. One can choose a particular $m_F$ ground state into which the atomic population is optically pumped beforehand. Therefore, this $m_F$ state can be used to represent the ground state, $\left|g\right>$, as for example in a Rydberg quantum repeater protocol~\cite{zhao2010efficient}. At the same time, the trapping scheme can be used in experiments where only specific $m_F$ states are used to represent atomic populations, as, for example, $\left|g\right>,\left|e\right>$ and $\left|s\right>$ in spin-wave atomic memories~\cite{heller2020cold}.

Meanwhile, one of the most promising applications of Rydberg states next to  waveguides is that of all-fibre based quantum networks, namely for quantum repeaters~\cite{zhao2010efficient}. Ideally, one would need to fully quantify the energy shifts of Rydberg states caused by the dielectric material as a function of the atom's distance from the waveguide. At the same time, for optimal performance of a Rydberg-based quantum repeater protocol,  trapping of the Rydberg atoms to reduce decoherence is also necessary. Both of these aspects could be achieved with our proposed trapping scheme. The proposed method for trapping atoms can also be applied to Cs atoms, provided  specific Rydberg states that meet the conditions for generating the trapping potential are identified.

\begin{acknowledgments} 
The authors would like to thank F. Le Kien, K. Mølmer, R. Löw, and M. Saffman for very insightful discussions and the Scientific Computing and Data Analysis Section at OIST. This work was supported in part by funding from  OIST Graduate University.  D.J.B. and S.N.C. acknowledge support from the Japan Society for the Promotion of Science (JSPS) Grant-in-Aid No.~22K13986 (Early Career) and No.~24K08289, respectively. D.F.K. acknowledges support from the Danish National Research Foundation through the Center of Excellence “CCQ” (Grant agreement no.: DNRF156).  E.B. acknowledges support from l’Agence Nationale de la Recherche (ANR), Project ANR-22-CE47-0011. 
\end{acknowledgments} 

\bibliography{NJP_arxiv.bib}

\end{document}